# Task-Parallelism in SWIFT for Heterogeneous Compute Architectures

Abouzied M. A. Nasar,[1★] Benedict D. Rogers,[1] Georgios Fourtakas,[1] Mladen Ivkovic,[2] Tobias Weinzierl,[2] Scott T. Kay,[3] and Matthieu Schaller[4,5]

[1] *School of Engineering. The University of Manchester, Nancy Rothwell Building, Manchester M13 9QS, UK*
[2] *Department of Computer Science, Durham University, South Road, Durham DH1 3LE, UK*
[3] *Jodrell Bank Centre for Astrophysics, Department of Physics and Astronomy, The University of Manchester, Manchester, M13 9PL, UK*
[4] *Lorentz Institute for theoretical physics, Leiden University, PO Box 9506, NL-2300 RA, Leiden, The Netherlands.*
[5] *Leiden observatory, Leiden University, PO Box 9513, NL-2300 RA, Leiden, The Netherlands.*

**ABSTRACT**

This paper highlights first steps towards enabling graphics processing unit (GPU) acceleration of the task-parallel smoothed particle hydrodynamics (SPH) solver SWIFT. Novel combinations of algorithms are presented, enabling SWIFT to function as a truly heterogeneous software leveraging task-parallelism on CPUs for memory-bound computations concurrently with GPUs for compute-bound computations while minimising the effects of CPU-GPU communication latency. The proposed algorithms are validated in extensive testing. The GPU acceleration methodology is shown to deliver up to ∼3.5 and ∼ 7.5 speedups for the offloaded computations when including and excluding the time required to prepare and post-process data transfers on the CPU side, respectively. The overall performance of the GPU-accelerated hydrodynamic solver for a full simulation on a single Grace-Hopper superchip is 1.8 times faster compared to the superchip's fully parallelised CPU capabilities. This constitutes an improvement from ∼8 million particle updates/s for the full CPU-only baseline (∼115'000 updates per CPU core) to ∼15 million updates/s for the GPU-accelerated SPH solver. Moreover, it displays near-perfect strong scaling on 4 Grace-Hopper nodes. The GPU-acceleration is also demonstrated to give a 29% improvement in energy efficiency in comparison to CPU-only baselines. Finally, inter-influential bottlenecks in the prototype solver presented in this work are identified: A significant amount of time (up to 80%) of a GPU-offloading cycle is spent on preparing and post-processing particle data on the CPU for the transfer to and from the GPU, respectively. Approaches are suggested to minimise their effects and maximise the solver's performance in our future work.

**Key words:** GPU acceleration – High-performance computing – Smoothed Particle Hydrodynamics – Software – Algorithms

## 1 INTRODUCTION

Numerical simulations are a pivotal tool in astrophysics and cosmology as they allow a rigorous interpretation of galaxy formation and evolution (e.g. Vogelsberger et al. 2019), planetary collisions (e.g. Kegerreis et al. 2019) and other complex phenomena (e.g. Andrade et al. 2021) and the validation of new and existing theoretical models for these phenomena. Due to the range of density and timescale variations and the fact that most of the universe's volume is highly rarefied, cosmology solvers often rely on Lagrangian methods, where resolution elements such as particles naturally move and cluster into relatively high density regions with the fluid flow. Examples of such Lagrangian methods are smoothed particle hydrodynamics (SPH), initially developed by Lucy (1977) and Gingold & Monaghan (1977). SPH is currently used in many state-of-the-art solvers (Wadsley et al. 2017; Price et al. 2018; Hopkins 2015b) alongside other Lagrangian methods such as moving mesh methods (Springel 2010; Chang et al. 2017) and finite volume particle methods (Hopkins 2015a). Conversely, Eulerian solvers such as Athena++ (Stone et al. 2020), PLUTO (Mignone et al. 2012) and RAMSES (Teyssier 2002) are also widely used. In contrast to Lagrangian methods, they typically rely on adaptive mesh-refinement (AMR) techniques to dynamically adapt the Eulerian mesh to ensure expensive high resolution is only used where required to capture the physics, thereby minimising computational effort. While SPH's intrinsic "adaptivity" doesn't require on-the-fly adaptive mesh reconstruction, its continually evolving unstructured particle connectivity challenges modern accelerators. It hence poses exciting algorithmic research questions, which can be generalised to a wider set of particle-based modelling techniques.

The scale of astrophysics simulations has been increasing rapidly and recently deployed digital twins of the universe, (e.g. Pakmor et al. 2023; Schaye et al. 2023; Frontiere et al. 2025), are allowing researchers to better understand the complexities of the tightly coupled non-linear physics which govern the universe's behaviour. Simulations at this scale require massive resolutions with $\mathcal{O}(10^{10})$ to $\mathcal{O}(10^{12})$ fluid elements (particles) in order to resolve the complex physics accurately and reproduce/resolve the observables for accurate comparison and quantification. Such large simulations must be run on supercomputers using codes designed to maximise computational scalability.

Given the recent advances in supercomputers, such as the Ex-aFLOP barrier having been broken (Atchley et al. 2023), and basing the argument on the increase in computational power available, one could argue that a step-change in our ability to resolve astrophysical phenomena using simulations is now within reach. However, the architecture of new exascale supercomputers which are benchmarked achieving $10^{18}$ floating point operations per second (FLOPS) is dominated by heterogeneous architectures employing graphics processing unit (GPU) acceleration. Hence, existing astrophysics solvers previously designed for massively parallel CPU execution must either be adapted to benefit from GPU acceleration or re-written entirely such that they are designed from the bottom-up to leverage GPUs.

★ E-mail: abouzied.nasar@manchester.ac.uk

Consequently, the development of general purpose CFD solvers capable of efficiently leveraging exascale architectures (especially current machines which are heterogeneous) is a very active field of research. Computational fluid dynamics (CFD) generally requires HPC for accurate flow simulation especially when attempting to accurately capture complex phenomena such as turbulence (López et al. 2020), combustion (Kolla & Chen 2018) and multi-phase flow (Karnakov et al. 2019). An extremely wide variety of approaches to developing exascalable CFD solvers have been proposed for different numerical techniques such as finite volume methods (Sharma et al. 2024; Antoniadis et al. 2022), mixed finite volume finite element methods (Gorobets & Bakhvalov 2022), spectral element methods (Merzari et al. 2024; Lindblad et al. 2024), finite difference methods (Deskos et al. 2020) and SPH (Guerrera et al. 2019). A methodology highly relevant to the research presented in this paper is the approach used by the Uintah solver of Holmen et al. (2022) which, whilst being mesh-based, also uses a tasking approach to distribute the workload between the available computational hardware. Uintah uses task schedulers to execute tasks simultaneously across the available hardware (CPUs and GPUs) based upon user-specification of which hardware to use for running different task types.

The conventional approach to GPGPU acceleration for SPH (Domínguez et al. 2022; Bilotta et al. 2016; Cercos-Pita 2015) is to have such data residing on GPUs for the entire simulation, as this allows the data to be re-used for all computations eliminating losses in efficiency induced by a von Neumann-esque CPU-GPU communication bottleneck. Massive gains in performance have been demonstrated by DualSPHysics (Domínguez et al. 2022), GPUSPH (Bilotta et al. 2016), SHAMROCK (David–Cléris et al. 2025), SPH-EXA (Cabezón et al. 2025), and HACC (Frontiere et al. 2025), which apply this method. This approach is especially advantageous for pre-exascale architectures where the GPU is connected to the compute node via a relatively slow PCIe port. Typically, CPU-GPU communications over conventional PCIe ports tend to be the first bottleneck encountered in many applications (Pabst et al. 2010; Xu et al. 2014). The importance of the CPU-GPU connection speed is hence widely recognised in the HPC community, which is evidenced at the time of writing by two of the world's leading supercomputers, Summit and Frontier, using AMD's Infinity Fabric and Nvidia's NVLink connections to deliver greater bandwidth, 72 GB/s (Pearson 2023) and 50 GB/s (Vergara Larrea et al. 2019) respectively, than via PCIe ports of the same generation (PCIe 4 and PCIe 3 with 32 GB/s and 16 GB/s, respectively).

At the same time, specialised HPC "superchips" such as Nvidia's Grace-Hopper (Schieffer et al. 2024) and AMD's MI300x (Smith et al. 2025) have been developed (partly due to the boom in artificial intelligence and machine learning). They are set to feature heavily in new supercomputer architectures (Herten et al. 2024; McIntosh-Smith et al. 2024; Fusco et al. 2024; Herten 2023) and offer large CPU memory (480 GB for the Grace CPU in the Grace-Hopper chip) complementing relatively smaller GPU memory (96 GB for its Hopper GPU) with CPU-GPU memory cache-coherency and a faster than conventional CPU-GPU connection, up to 7x faster than current PCIe 5 ports for the GH200 "superchip" (Fusco et al. 2024)). This allows for the usage of larger data sets than possible with the conventional general-purpose GPU (GPGPU) computer architectures and minimises loss in GPU acceleration efficiency due to slow CPU-GPU data transfers. New pre-exascale machines, such as the European supercomputer Jupiter (Herten et al. 2024) and the UK based Isambard-AI (McIntosh-Smith et al. 2024), leverage these "superchips" designed specifically to overcome the limitations associated with the smaller GPU memory. As such, a question presents itself as to whether a GPU-acceleration approach which *doesn't* store particle data permanently in GPU memory may be worthwhile on newer hardware. As CPU memory is typically larger than GPU memory on a node, this would permit solving (much) larger problems on the same hardware.

Parallelisation of SPH on single compute nodes using shared memory parallelisation strategies with OpenMP or Threading Building Blocks (TBB) for example is relatively straightforward and has been applied effectively in many SPH solvers for engineering and other terrestrial applications, see (e.g. Ramachandran et al. 2021; Zhang et al. 2021). However, the scale of astrophysics and cosmology applications requires extreme parallelisation on thousands of compute nodes and as such, advanced extremely scalable techniques for distributed memory parallelisation are required. These strategies must overcome a number of obstacles, the largest of which being that when the problem is scaled, adequate distribution of the computational workload across potentially heterogeneous resources with heterogeneous capabilities becomes difficult. The Lagrangian nature of SPH leads to the computational points (particles) leaving and entering different parts of the distributed memory, adding to this difficulty and necessitating robust domain decomposition strategies and dynamic load balancing (Guo et al. 2018). Temporal adaptivity as required for feasible time-to-solution where particles have different time step sizes depending on their local conditions such as their interaction radius and CFL condition can also have a strong influence on the computational work available to each node at a given point in the simulation (Borrow et al. 2018), further complicating load-balanced domain decomposition.

In cases involving gravitational forces, particles cluster in regions of the domain which again complicates load balancing and domain decomposition and techniques such as octree-based decomposition become a necessity (Springel 2005; Schaller et al. 2024; Menon et al. 2015). However, octree-based methods are difficult to implement efficiently in GPU solvers as they require irregular memory access patterns and involve complex data dependencies, but there are efforts to do so (Keller et al. 2023). Other approaches attempt to avoid this limitation of octrees by using binary radix trees for example (David–Cléris et al. 2025), which simplify and minimise the data access patterns. Tree linearisation using space filling curves and breadth-first tree traversal (Weinzierl 2019) can also improve concurrency in tree-based methods and in-turn their efficiency on GPUs (Garanzha et al. 2011). The requirement of solving other physics such as chemical and nuclear reactions and radiation for example, only in localised regions in the domain, complicates load balancing even further for astrophysics applications.

Our work focuses on the open-source astrophysics solver SWIFT[1] (Schaller et al. 2024), which uses task-parallelism instead of conventional data-parallelism to maximise the solver's scalability, especially for complex simulations involving large scale disparity requiring ex-treme spatial and temporal adaptivity (Borrow et al. 2018). SWIFT relies on QuickSched (Gonnet et al. 2016) which defines all the computations and communications required for a simulation as small fine-grained tasks. These can be executed by any of the available computational threads as and when their dependencies (completion of other tasks) have been fulfilled. The task-parallelism approach minimises idle time due to load imbalance and in principle allows for MPI communications to be interleaved with computations such that data transfer times between MPI ranks are hidden underneath computations. SWIFT was shown to scale exceptionally well on CPUs (with only 15% loss in performance when weak scaling from one compute node with 128 CPU cores to 343 compute nodes with more than 40k CPU cores) (Schaller et al. 2024). Aiming to benefit from the performance and scaling properties facilitated by SWIFT's task-parallel approach, the goal of this work is to develop a prototype of a GPU-accelerated hydrodynamics solver while retaining the task-parallelism.

[1] swift.strw.leidenuniv.nl

However, GPU-acceleration of the task management is a difficult endeavour as task management is dominated by sequential and atomic accesses and logic operations. It requires random access to many different small packets of data. Moreover, the cosmology problems SWIFT is designed for require fine-grained tasks with irregular workloads and numerous inter-dependencies, which exacerbates the complexities of designing an efficient GPU-managed task scheduler. As such, the development of an efficient GPU-acceleration of a task scheduler is difficult (Chalk 2017) and remains an open problem (e.g. Morgenstern 2024).

Instead, in this work we explore retaining the task scheduling on the CPU while offloading the (computationally intensive) tasks themselves to the GPU. More precisely, task management, tree construction and traversal and other highly memory bound tasks are left for the CPU, while compute intensive but less memory bound hydrodynamics tasks are accelerated by the GPU. Moreover, aiming to benefit from the larger CPU memory (compared to GPU memory) allowing the solution of bigger problems per node used, our approach is to also keep particle data in CPU memory and to transfer it as needed for specific computations.

In this paper, we present this new implementation of a GPU-accelerated task-parallel SPH solver in SWIFT and we address some of the challenges faced whilst developing such a solver. The new implementation capitalises on over-lapping CPU-GPU communications with GPU computations. Tasks are bundled into large groups such that the CPU-GPU communications are optimised for high bandwidth and kernel launch overheads are minimised. We also identify and highlight the challenges associated with exploiting GPU hardware in heterogeneous systems and outline pathways to alleviate currently present bottlenecks in future work.

This paper is structured as follows; in Section 2, the methodology used in the GPU-accelerated SPH solver is presented, whereas the SPH method is outlined in Section 2.1. Our task-parallel implementation is discussed in Section 2.2 and the method for GPU-acceleration is presented in Section 2.3. In Section 3, the speedups achieved for task computations are presented. Three different systems built with different hardware architectures are used for the analysis. In Section 4, the GPU-accelerated solver's performance for full simulations is assessed and compared to the CPU-only SWIFT solver. Current limitations of the GPU-accelerated solver are also discussed in Section 4 before conclusions and future work are discussed in Section 5.

## 2 METHODOLOGY

### 2.1 The Governing Equations

#### *2.1.1 The SPH interpolation*

Within the SPH scheme, the computational domain is discretised into Lagrangian particles which move through space according to the governing equations. A quantity $\mathbf{Q}$ of a particle $i$ may be interpolated as

$$\hat{\mathbf{Q}}_i = \sum_j \frac{m_j}{\rho_j} \mathbf{Q}_j W_{ij} \tag{1}$$

where $W$ is the smoothing kernel with $W_{ij} = W(r_{ij}, h_i)$, $r_{ij} = |\mathbf{r}_j - \mathbf{r}_i|$ with $\mathbf{r}$ being the position vector, $m$ is mass of the particle, $\rho$ is the mass density, $h$ is the smoothing length and $j$ represents the neighbouring particles. Examples of widely used kernels $W$ are the cubic spline and Wendland kernels.

In SWIFT, $h$ is variable and is a particle carried quantity. For SPH with uniform $h$, neighbouring particles $j$ are included in the summation if the distance between them and particle $i$ is smaller than $H = \Gamma_{\mathrm{K}} h$, where $H$ is the kernel's cut-off radius or compact support radius, and $\Gamma_{\mathrm{K}}$ is a kernel dependent constant (see Dehnen & Aly 2012). To increase the number of interacting neighbouring particles, the smoothing length of the kernel may be adjusted by a constant $\eta$ defined such that

$$\hat{n}(h) \left(\frac{h}{\eta}\right)^{n_d} = 1 \tag{2}$$

where $\hat{n}$ is the particle number density and $n_d$ is the number of spatial dimensions. $\eta$ is usually set to between 1 and 2. For applications with variable $h$, when a particle $a$ with a smaller $h$ is a neighbour of a particle $b$ due to $b$ having a larger $h$, particle $b$ is also a neighbour (included in the summation) of particle $a$ to ensure momentum and energy conservation. In typical SPH applications which require variable $h$, particle densities vary greatly, and a particle's density is directly linked to its smoothing length. As such, $h$ is evaluated individually for each particle following the process described in Section 2.1.2.

From Eq. 1, the fluid density may be interpolated as,

$$\hat{\rho}_i = \sum_j m_j W_{ij}\ . \tag{3}$$

Similarly to the interpolation of a function in Eq. 3, its spatial derivatives may be approximated as

$$\nabla \cdot \hat{\mathbf{Q}}_i = \frac{1}{\rho_i} \sum_j m_j \mathbf{Q}_j \cdot \nabla W_{ij}\ , \tag{4}$$

$$\nabla \times \hat{\mathbf{Q}}_i = \frac{1}{\rho_i} \sum_j m_j \mathbf{Q}_j \times \nabla W_{ij}\ . \tag{5}$$

SWIFT offers a range of approaches for modelling hydrodynamics using these SPH interpolations such as the "minimal" formulation based on Price (2012) and the GADGET 2 formulation based on Springel (2005). The default recommendation is to use the SPHENIX formulation (Borrow et al. 2022) which is more suited to SWIFT's major use-cases (galaxy formation applications). As such, for GPU acceleration the focus is on the most computationally intense aspects of the SPHENIX scheme, namely the "density", "force" and "gradient" loops which solve the continuity and momentum equations. Full details of the CPU solver implementation in SWIFT can be found in Schaller et al. (2024). The GPU-accelerated parts of the solver are summarised in the following sub-sections.

### 2.1.2 The Density Loop

In SWIFT, the variable smoothing length and density of a particle are coupled via spatial interpolations which solve for them iteratively such that an estimate of the smoothing length is arrived at which ensures all particles interact with a pre-defined weighted number of neighbouring particles (Price 2012). This ensures the SPH interactions are accurate but also computationally efficient. First, the particle number density $n$ and its gradient with respect to the smoothing length are evaluated for all particles in the domain via

$$\hat{n}_i = \sum_j W_{ij} \,, \tag{6}$$

$$\frac{\mathrm{d}\hat{n}_i}{\mathrm{d}h} = -\sum_j \left( \frac{n_d}{h_i} W_{ij} + \frac{r_{ij}}{h_i} \nabla_i W_{ij} \right) \,, \tag{7}$$

where $n_d$ is the number of spatial dimensions for the interpolation.

The mass density $\rho$ is then calculated using Eq. 3 and a Newton-Raphson scheme is used to iteratively solve for Eq. 6 and Eq. 7 until a relative tolerance of $10^{-4}$ is reached. The "density" computations in SWIFT only involve the initial interpolations in Eq. 6 and Eq. 7, whilst the iterative solution to find the optimal smoothing length is performed in another "ghost" computation. In our GPU-accelerated solver, we perform the ghost computation on the CPU, as it is only compute intensive for the very first time step, when the predicted and expected densities are disparate. In that case, a few (less than 10) iterations are required for convergence. For the remainder of the simulation, the time required to compute these ghost tasks becomes insignificant in comparison to the main SPH loops. N.B. in very rare cases, the particle spatial distribution may become very poor in the vicinity of a handful of particles such that convergence requires $\gg 10$ iterations for these particles. However, the task-based nature of the code allows other tasks to be executed for the vast majority of the simulation domain such that these iterations only keep a handful of CPU threads busy whilst all other threads are occupied with other task subtypes. As such, retaining the ghost computations on the CPU is not a significant performance concern.

### 2.1.3 The Gradient Loop

In this second particle interaction loop, the dissipative terms in the momentum equation are evaluated using an artificial viscosity formulation based on Cullen & Dehnen (2010) which follows on from the approach of Morris & Monaghan (1997). The artificial viscosity is a particle-carried quantity evolved through time in a manner which decays when a particle is not near shock regions. The viscous term reads

$$\left.\frac{\mathrm{d}\mathbf{v}_i}{\mathrm{d}t}\right|_{\rm visc} = -\sum_j m_j \frac{\nu_{ij}}{2} \left( f_i \nabla_i W_{ij} + f_j \nabla_j W_{ij} \right) \tag{8}$$

where $f$ is the $h$-factor which accounts for non-uniform $h$ written as

$$f_i = 1 + \frac{h_i}{n_d \hat{\rho}_i} \frac{\mathrm{d}\hat{\rho}_i}{\mathrm{d}h} \,, \tag{9}$$

$$\nu_{ij} = \frac{\alpha_{\nu,ij}\, \mu_{ij}\, v_{\mathrm{sig},ij}}{\hat{\rho}_i \hat{\rho}_j} \,, \tag{10}$$

and the signal velocity is written as

$$v_{\mathrm{sig},ij} = c_{s,i} + c_{s,j} - \beta \mu_{ij} \,, \tag{11}$$

with

$$\mu_{ij} = \begin{cases} \frac{\mathbf{v}_{ij} \cdot \mathbf{r}_{ij}}{|\mathbf{r}_{ij}|} & \text{if } \mathbf{v}_{ij} \cdot \mathbf{r}_{ij} < 0 \\ 0 & \text{otherwise} \end{cases} \,, \tag{12}$$

where the constant $\beta = 3$. The artificial viscosity parameter $\alpha_\nu$ is evolved in time via

$$\alpha_{\nu,i}(t + \Delta t) = \alpha_{\nu,i}(t) - \alpha_{\nu,i,\mathrm{loc}} \,, \tag{13}$$

$$\alpha_{\nu,i,\mathrm{loc}} = \alpha_{\nu,\max} \left( \frac{S_i}{v^2_{\mathrm{sig},ij} + S_i} \Delta t \right) \,, \tag{14}$$

$$S_i = H_i^2 \max\left(0, -\dot{\nabla} \cdot \mathbf{v}_i\right) \,, \tag{15}$$

$$\dot{\nabla} \cdot \mathbf{v}_i = \frac{\nabla \cdot \mathbf{v}_i(t + \Delta t) - \nabla \cdot \mathbf{v}_i(t)}{\Delta t} \,. \tag{16}$$

where $H_i$ is the kernel cut-off radius.

The final value of the artificial viscosity term is then evaluated as

$$\alpha_{\nu,ij} = \frac{\alpha_{\nu,i} + \alpha_{\nu,j}}{2} \frac{B_i + B_j}{2} \tag{17}$$

where $B$ is the Balsara switch (Balsara 1990)

$$B_i = \frac{|\nabla \cdot \mathbf{v}_i|}{|\nabla \cdot \mathbf{v}_i| + |\nabla \times \mathbf{v}_i| + \varepsilon c_{s_i}/h_i} \,, \tag{18}$$

$c_{s,i}$ is the local speed of sound and $\varepsilon = 0.0001$ is a dimensionless constant to prevent division by zero.

### 2.1.4 The Force Loop

In the force loop, the acceleration due to pressure gradients and the rate of change of energy are evaluated using

$$\frac{\mathrm{d}\mathbf{v}_i}{\mathrm{d}t} = -\sum_j m_j \left( \frac{f_i P_i}{\rho_i^2} \nabla_i W_{ij} + \frac{f_j P_j}{\rho_j^2} \nabla_i W_{ij} \right) \,, \tag{19}$$

$$\frac{\mathrm{d}u}{\mathrm{d}t} = \sum_j m_j \frac{f_i P_i}{\rho_i^2} \mathbf{v}_{ij} \cdot \nabla_j W_{ij} \,, \tag{20}$$

where $u$ is internal energy and $\nabla_i$ and $\nabla_j$ are the gradients centred at particles $i$ and $j$ respectively.

Artificial thermal conduction is also used in the SPHENIX formulation (Borrow et al. 2022), which is added into Eq. 20 as

$$\left.\frac{\mathrm{d}u}{\mathrm{d}t}\right|_{\rm diff} = \sum_j \alpha_{c,ij} v_{c,ij} m_j (u_i - u_j) \frac{f_{ij} \nabla_i W_{ij} + f_{ij} \nabla_j W_{ij}}{\rho_i + \rho_j} \tag{21}$$

with the artificial conduction strength $\alpha_{c,ij}$ given by a pressure weighting of the contribution of each interacting particle

$$\alpha_{c,ij} = \frac{P_i \alpha_{c,i} + P_j \alpha_{c,j}}{P_i + P_j} \tag{22}$$

and is evolved through time by integrating its rate of change which is given by

$$\frac{\mathrm{d}\alpha_{c,i}}{\mathrm{d}t} = \beta_c H_i \frac{\nabla^2 u_i \alpha_{c,i}}{\sqrt{u_i}} - (\alpha_{c,i} - \alpha_{c,i,\min}) \frac{v_{c,i}}{H_i} \,, \tag{23}$$

$$v_{c,ij} = \frac{|\mathbf{v}_i \cdot \mathbf{r}_{ij}|}{|\mathbf{r}_{ij}|} + \sqrt{2 \frac{|P_i - P_j|}{\rho_i + \rho_j}} \,, \tag{24}$$

where $0 < \alpha_{c,i,\min} < 2$ and $\beta_c = 1$ (a non-dimensionless parameter). Also, conduction is limited in regions with strong shocks by limiting $\alpha_c$ using

$$\alpha_{c,\max,i} = \alpha_{c,\max} \left( 1 - \frac{\alpha_{\nu,\max,i}}{\alpha_{\nu,\max}} \right) \,, \tag{25}$$

with the constant $\alpha_{c,\max} = 1$ and then setting $\alpha_c$ to

$$\alpha_{c,i} = \begin{cases} \alpha_{c,i} & \alpha_{c,i} < \alpha_{c,\max} \\ \alpha_{c,\max} & \alpha_{c,i} \geq \alpha_{c,\max} \end{cases} \,. \tag{26}$$

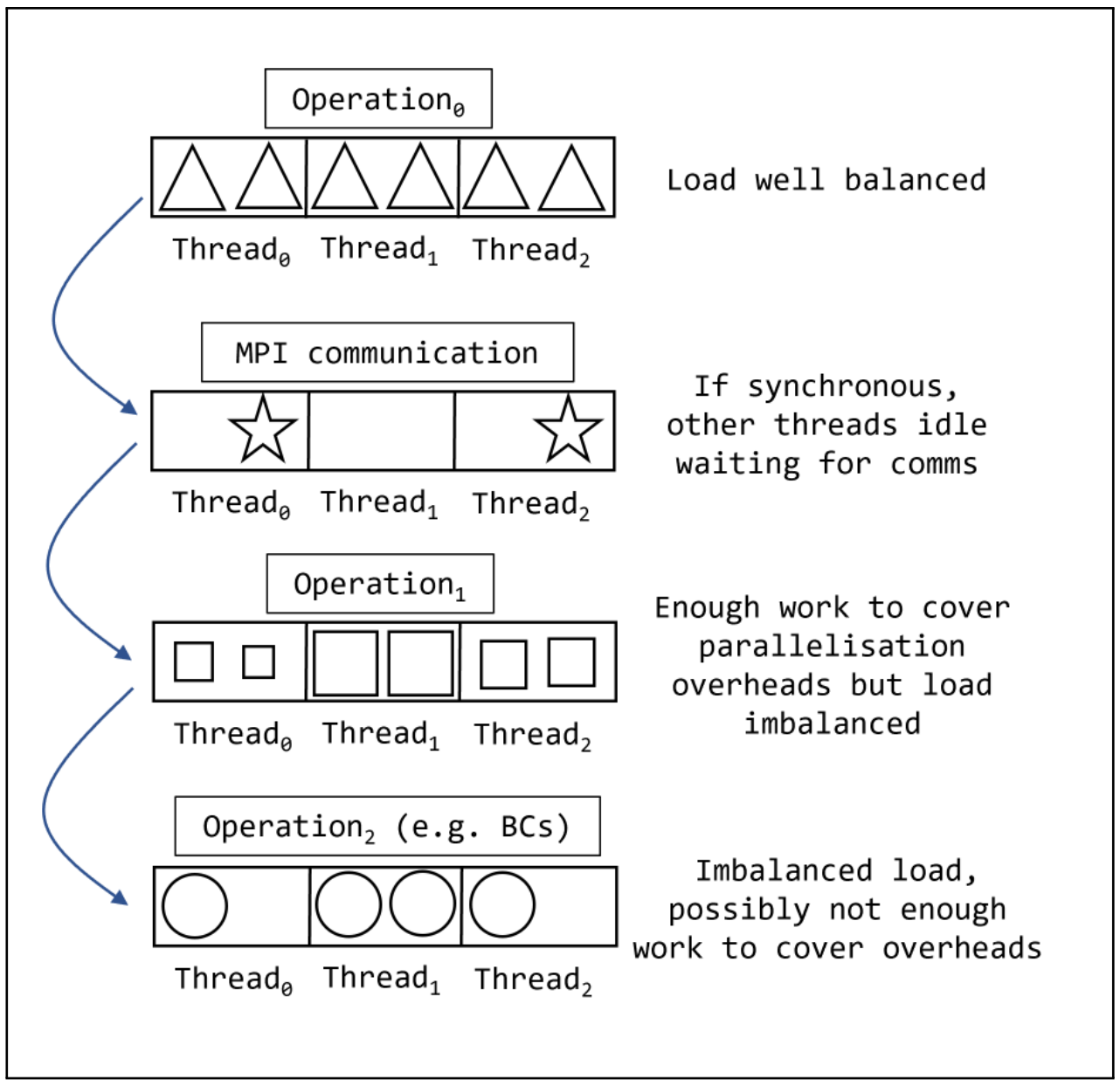

**Figure 1.** Illustration of data parallelisation where all threads operate on pieces of data performing the same operation concurrently and the computation is performed one process at a time. The problem of enforcing boundary conditions (BCs) is used as an example in Operation$_2$.

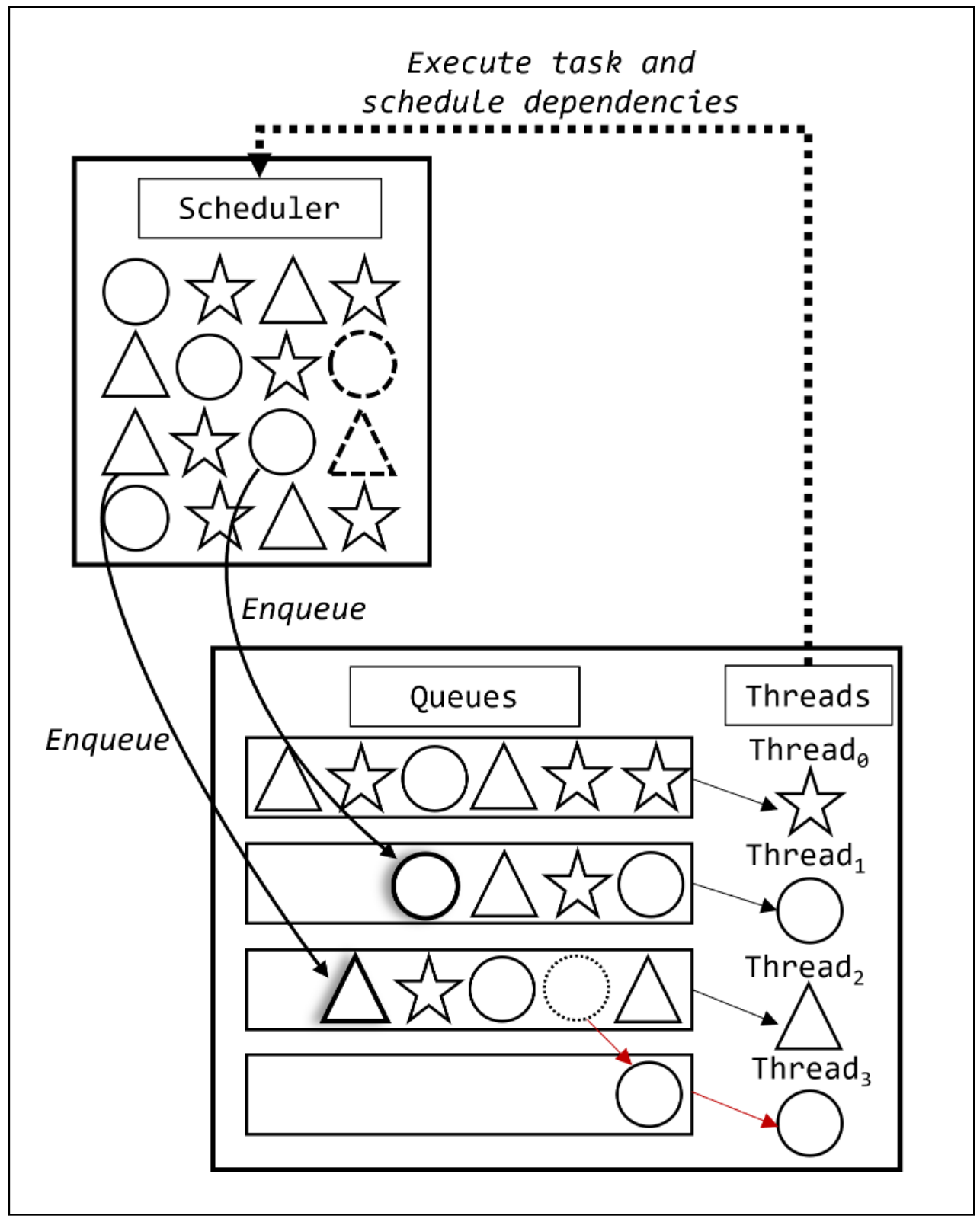

**Figure 2.** Illustration of task parallelism where a scheduler pools tasks from different processes and enqueues them in queues for all the available threads. Threads execute tasks from different processes concurrently, then unlock their dependent tasks before moving onto the next task in their queue. Threads can also steal tasks from other queues as illustrated by Thread$_3$ stealing the circular task from Thread$_2$.

### 2.2 Task-Parallelism Within SWIFT

In SWIFT simulations, the computational domain is split into cubic top-level cells, which form the roots of a cell tree, each of which contains a set of SPH particles. A recursive process is used to traverse the cell-tree to find daughter cells (daughters of the top-level cells) containing $O(10^3)$ particles. This target number of particles may not be achievable for rare cases where the particles in a daughter cell have compact support radii $H$ which are larger than the cell width. In such cases, cells are chosen at a level with widths larger than the maximum $H$ of particles in the cell, $H_{max}$. These daughter cells, containing $O(10^3)$ particles, are the cells on which tasks, discussed below, are created. Note, however, that the cell tree is still refined further to facilitate fast particle-particle interaction computations. The leaf cells of the tree typically contain $O(10^1) - O(10^2)$ particles (in 3D) and a recursive mechanism discussed in Sections 2.2.2 and 2.3.1 is used to traverse the tree to a level where the cell width is just larger than $H_{max}$.

The domain-wide computations (inter-particle interactions, particle-wise operations such as time integration, etc.) required for the simulation are decomposed into sets of relatively small tasks. For inter-particle interaction computations, each task is a set of computations performed either on a single task-level cell (discussed above) or for particles in a pair of adjacent task-level cells. By construction of the cell widths to fit $H_{max}$, this approach guarantees all particles' neighbours would be found. It furthermore allows for computational parallelisation to be implemented such that each computational thread can perform an individual task, which is the essence of task-parallelism.

#### *2.2.1 Task-Parallelism*

With conventional data-parallelism, computations are performed in a manner where multiple threads perform the same operation on different data elements. Commonly, this requires explicit synchronisation points between different operation types. For example, each operation type required for the simulation, e.g. computation of particle accelerations, time integration, and MPI communications, must be completed before operations of another type may be executed, as illustrated in Figure 1. This can detriment performance in cases where the computational load is either imbalanced or insufficient: The fundamental principle of data parallelism is to have multiple threads perform *the same* operations on multiple data. If for any reason some threads have less data to work on then others and run out of data to work on, they will wait idly at a subsequent synchronisation point. Similarly, if the performed operation has an a-priori unknown duration such as a method iterating until a tolerance threshold is met, some threads may complete their work faster than others, also leading to idle time at synchronisation points.

While these inefficiencies may be alleviated through use of more sophisticated scheduling techniques such as dynamic scheduling (as available in OpenMP and TBB for example). Task-parallelism attempts to eliminate such load imbalances by using a fundamentally different approach: Threads work concurrently on *both different data and different operations*. Threads which have completed their task (unit of work on a set of data) can move on to a next task, as provided in a queue by a task scheduler, independently of which operations other threads are busy with, as shown in Figure 2. Note that dynamic scheduling in data-parallel APIs introduces additional overheads which may be significant depending on the size of the problem (Kasielke et al. 2019). The same is true for task management overheads in task-based methods and we discuss the necessary considerations in Section 2.2.2. Figure 1 illustrates examples where maintaining efficiency with data-parallel approaches is a complex undertaking.

In a naïve task-parallel implementation, task queues could be filled unequally as shown in Figure 2. However, in SWIFT a thread with a relatively empty queue "steals" tasks from other over-filled queues (for example Thread$_3$ in Figure 2 has stolen a task from Thread$_2$ as its queue was empty). Task stealing helps ensure near-perfect load balancing (Schaller et al. 2024) and is a commonly used optimisation in task-parallel methods (McIntosh-Smith et al. 2023).

As shown in Figure 2, with task-parallelism, one compute thread may be computing accelerations for particles in one cell (triangle shaped task) whilst another thread may be integrating particles in another cell through time (circle shaped task) in tandem with another thread interleaving MPI communications with these computations.

For operations which are only required in small areas of the domain, e.g. wall boundary conditions, task-parallelism can significantly improve scalability and reduce thread idle time. These relatively small computations can be pooled with tasks from other operations to ensure compute threads always have a large pool of tasks available. This is used effectively in the Uintah solver which leverages task-parallelism to efficiently solve multi-physics problems (Meng & Berzins 2014) requiring multiple solvers with different discretisations and compute loads. Finally, communications are effortlessly made asynchronous with task-parallelism: they may be initiated by one thread asynchronously which can then move onto another compute task whilst the communication is finalised by the network controller.

In SWIFT, the correct order of operations is upheld through the implementation of dependencies in-between different task types. These dependencies ensure dependent tasks are only scheduled for execution after their dependency tasks are executed, as shown in Figure 2. SWIFT furthermore implements inter-task conflicts to avoid data races: While dependencies guarantee the correct order of operations, conflicts ensure that no two threads can act upon the same piece of data simultaneously *without* the use of atomic operations in performance critical sections. This is realised by lock mechanisms, whereby a thread executing a task obtains and holds a lock on the task's data until the task is complete. Details of these mechanisms may be found in Gonnet et al. (2016), where SWIFT's task engine, QuickSched, is also discussed in detail.

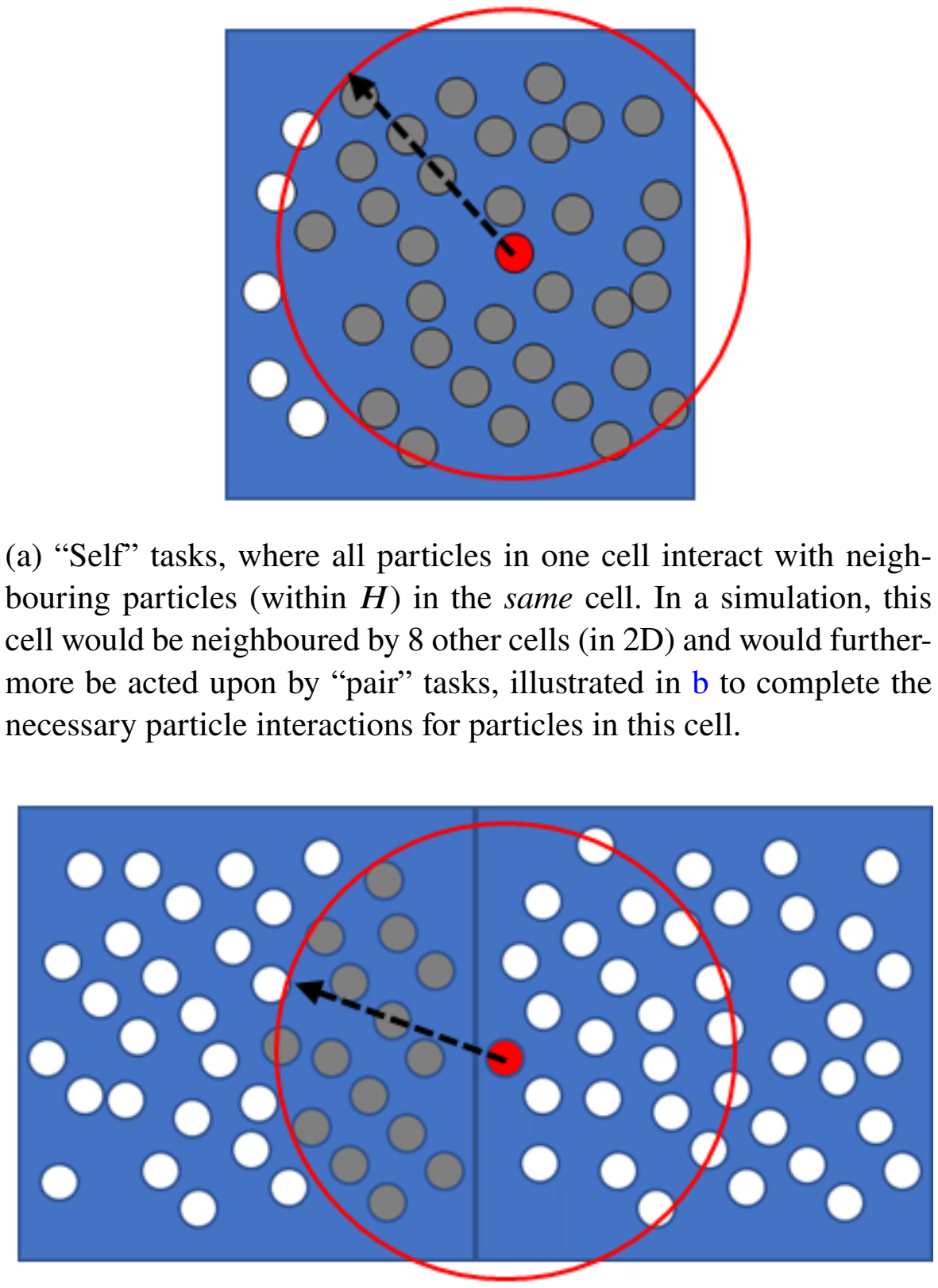

(a) "Self" tasks, where all particles in one cell interact with neighbouring particles (within $H$) in the *same* cell. In a simulation, this cell would be neighboured by 8 other cells (in 2D) and would furthermore be acted upon by "pair" tasks, illustrated in b to complete the necessary particle interactions for particles in this cell.

(b) "Pair" tasks, where all particles in one cell interact with neighbouring particles in an *adjacent* cell within their kernel support, and vice-versa. This is one of 8 possible pair tasks involving the cell of interest (in 2D). The number of pair tasks involving the cell of interest would be 26 in 3D.

**Figure 3.** SWIFT SPH task types.

### 2.2.2 *The SWIFT SPH tasks*

In SWIFT, tasks are categorised into types and sub-types. The major task types are "self" and "pair" tasks, which refer to the particle-particle interaction operations performed between particles of a single cell or between particles of two adjacent cells, respectively, as illustrated in Figure 3.

The SPHENIX scheme requires three sequential particle-particle interaction loops, as outlined in Section 2.1. Within the tasking system, they are modelled as three distinct "task-subtypes" consisting of a collection of self and pair tasks and are named "density", "gradient", and "force", respectively. In general, they are SPH summations which perform computationally intensive but memory-bound interactions between particles.

Other tasks include the "kick" and "drift" tasks for time integration and other "ghost" tasks which finalise the density, gradient and force tasks by performing minor particle-wise computations not involving neighbours. The aforementioned operations are of low computational intensity and are $O(N_t)$, where $N_t$ is the total number of particles. In our offloading approach, the data for these low intensity operations resides on CPU memory, as is the data for all operations. As a consequence, the path of access to their data for CPU threads is much shorter than the path for the GPU threads. As such, the offloading of these tasks is inefficient. Instead, they can be executed rapidly on CPU threads whilst GPU threads are busy executing the compute intensive density, gradient and force tasks, which are $O(N_c^2)$ where $N_c$ here is the number of particles in the cells used to compute the inter-particle interactions.

The overheads of task creation and management could be significant if tasks are too fine-grained. Thus, "tasking" overheads (task scheduling and activation, data retrieval, etc.) must be minimised to allow the solver to achieve the desired gains in time to solution. As such, SWIFT's tasks are designed to perform the particle-wise computations for all particles in a large parent cell (see Figure 4) which contains several daughter cells of the optimal size for SPH computations (just larger than $2H_{max}$). A CPU thread recursively loops through a task's daughter cells on a pre-defined cell tree, whereby the necessary self and pair interactions (with itself and other neighbouring "sister" cells) are performed for each leaf cell at the level optimal for performance of the core SPH computation. Attaching tasks to cells higher in the tree hierarchy in this manner, rather than at the leaf level, means that they are attached at an optimal level to minimise scheduling overheads. This recursion also allows for dynamic adaptation of the tasks to enable the large variation of particle smoothing lengths. It furthermore ensures that the evolution of dense and rarefied regions which occur in astrophysics simulations is handled accurately (all necessary interactions are performed) and as efficiently as possible. Note that the recursion process moreover permits the skipping of unnecessary interactions: no computations are attempted for pairs of leaf cells not connected by a face, edge or vertex, nor for cells which only contain particles that are not being integrated in time during the current simulation step.

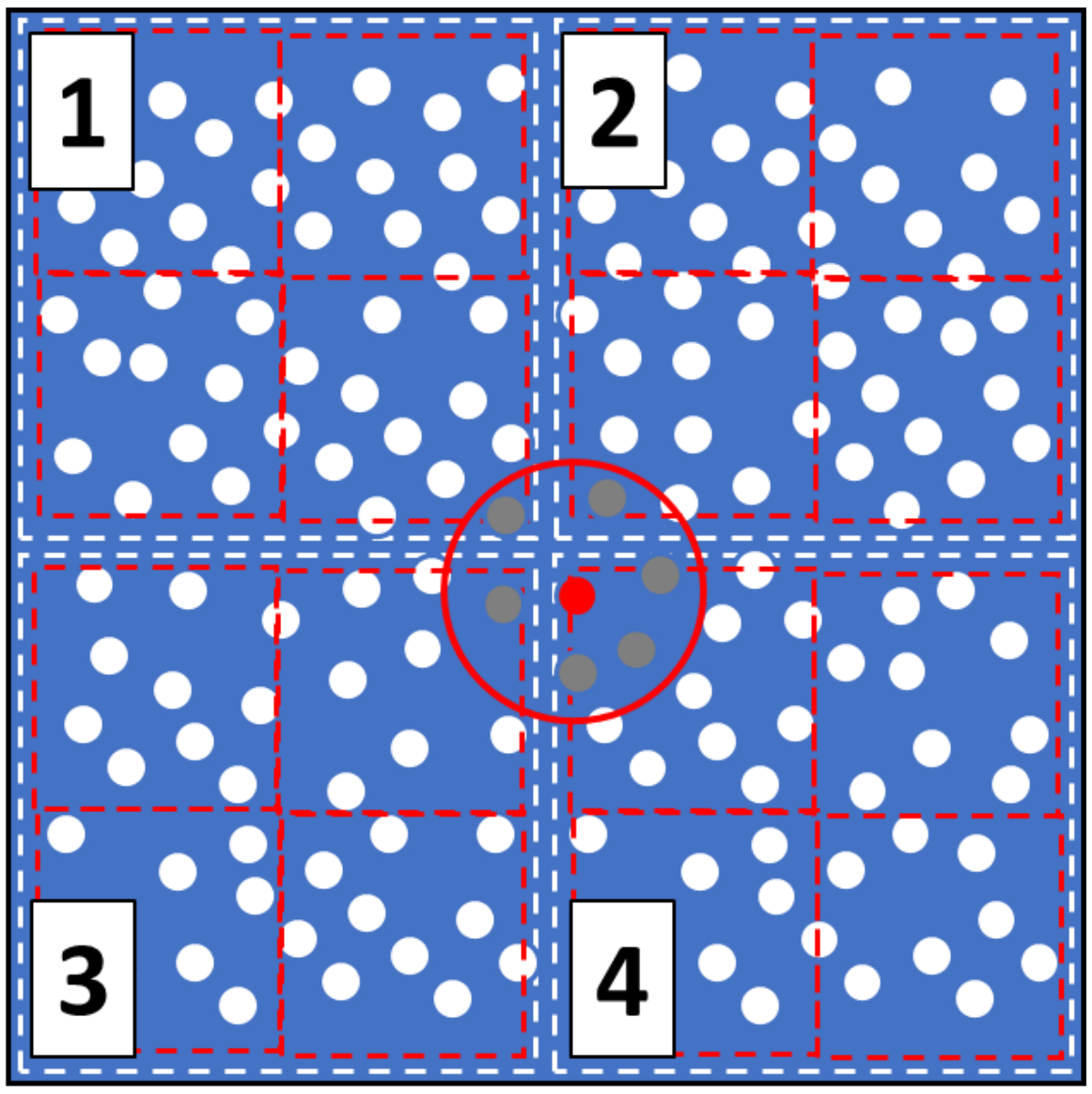

**Figure 4.** Recursive tasks. In SWIFT, tasks are usually created to act on a cell at a high level of the cell tree. As the high-level "parent" cell contains too many particles for efficient neighbour finding, the task recurses through the cell's daughter cells (4 cells in 2D, outlined in dashed white lines). If the cell's daughter cells are still too large for efficient particle interactions, SWIFT recurses until the width of the cell is just greater than the largest $H$ in the cell. At this point, SWIFT computes the resulting "self" and "pair" interactions using the optimum number of neighbours. In the example illustrated in this Figure, two recursions were executed: The first recursion split the cell into cells 1, 2, 3 and 4 (outlined in dashed white lines). However, the width of cells 1 to 4 is still twice greater than $H_{max}$ for the particles they contain. Cells 1 to 4 are therefore split into 4 more daughter cells each (outlined in dashed red lines). At this point, the cell width is just greater than the $H_{max}$ and the cells are not split further. In the case illustrated, particles are now contained in daughter cells 2 levels down the cell tree. Their "self" and "pair" type interactions are performed at this level in the tree. Using the red particle to illustrate, the red particle will only search for neighbours in adjacent daughter cells connected by a face or a vertex in 2D (dashed red) and will only interact with particles in those adjacent cells within its kernel support. In 3D, adjacent cells would either share a vertex, edge or face.

The cell tree is created at the beginning of the simulation and re-created when the particle distribution has changed significantly according to criteria discussed in Gonnet et al. (2013). Very coarse top level cells are defined once at the beginning of the simulation and are used as roots from which all daughter cells are spawned. In rare cases where the computational domain expands beyond these cells or cases where the particles grow to be as large as the top level cells, the top level cells are re-defined accordingly, on-the-fly, and the cell tree is re-created. During cell tree creation, the code recurses through a few top-level cells splitting them into daughter cells and organising particles into the daughters at each level created. Recursive cell creation continues until the level just above the level where daughter cells contain no particles, at which point the recursion stops. This procedure is different from the recursion within the cell to which a task is attached, highlighted in Figure 4. In the recursion during cell tree creation, the aim is to define the cell tree at deep enough levels which ensures that there is scope for cells to contain smaller particles if substantial particle migration occurs before the cell tree is re-created.

Additionally, for all pair interactions a pre-sorting step (Willis et al. 2018) is conducted prior to computing the particle interactions. Particles in two adjacent cells are sorted along the axis connecting the cells. This sorted ordering is stored in lists. It is used to reduce the number of distance checks between particles in pair task interactions by a significant amount: rather than checking all particles of the adjacent cell for their distance, the sorted lists are traversed. The sorting along the axis provides a lower boundary on the distance between two particles and enables the neighbour loop to stop early, discarding particles which are now known to be out of range. The pre-sorting step is a pseudo-Verlet list algorithm as detailed in Gonnet et al. (2013). Further details of the cell splitting and sub-task mechanisms are also discussed in Gonnet et al. (2013).

## 2.3 GPU acceleration

GPU acceleration is extremely beneficial for highly parallelisable and compute-bound problems. In the following sub-sections, different approaches are discussed for capitalising on the capabilities of GPUs for SPH simulations.

### *2.3.1 GPU acceleration for SWIFT*

SWIFT's SPH tasks are designed to be fine-grained to maximise efficiency by minimising thread idle time. All available compute threads can fetch tasks from their queues until all tasks are completed. The optimal size of these tasks is problem dependent, but typically they act on $O(10^3)$ particles. This task size is efficient for parallelisation on CPUs as tasks fit well on CPU cache (especially if the recursive mechanism is employed to identify daughter tasks acting on $O(10^2)$ particles) and memory access of small data packets is generally much faster on CPUs than it is on GPUs. However, in order to ensure efficient GPU off-loading of SWIFT's tasks, an approach is required which can maximise the efficiency of:

(i) *CPU-GPU communications*. It is much more efficient to transfer one large packet of data stored in simple 1D arrays than it is to transfer multiple small packets of data stored in complex structures, as this allows the interconnect to achieve much higher transfer speeds and minimises latency.

(ii) *Data transfer from global GPU memory to the GPU registers.* This is much more efficient when large packets of contiguous data are transferred, as this minimises memory transactions and helps saturate device memory (DRAM) bandwidth, improving transfer speeds.

(iii) *Computations on the GPU*. It is also more efficient to instruct the GPU to launch one kernel acting on a large block of data than it is to launch several kernels acting on smaller packets of data. This minimises kernel launch overheads and also benefits DRAM bandwidth saturation which is a major concern with GPU programming.

NVIDIA corporation (2025) provide a guide where these and other performance optimisation requirements are discussed in detail.

### *2.3.2 Heterogeneous SWIFT With Task-Parallelism*

Not all of SWIFT's tasks would benefit from GPU acceleration. With the concerns raised above in mind, a bespoke approach is proposed to off-load SWIFT's tasks to the GPU as follows.

**Host-side data packing**

On the SWIFT host code, threads performing SPH tasks access the particle data on which they will act by first de-referencing a pointer to find the location in memory of a "cell" struct corresponding to the cell which contains the particles. Host-code data is typically stored as structs of arrays of structs (SoAoS), where the cell struct contains an array of "part" structs, each of which contains scalar and $3 \times 1$ vector variables representing particle properties. Transferring data of this form directly to the GPU is intractable and requires nested CPU to GPU copies: first for the particle structs, and then for the cell struct which contains the particle structs. Assuming the data for $M$ cells must be sent to the GPU, each containing $m$ particles, first a copy instruction must be issued to transfer $M$ cell structs to the GPU. Then, an additional $M$ copy instructions must be issued to transfer the particle data into these GPU-side cell structs. If this procedure had been implemented, then host-side data manipulation, packing it from it's original CPU form to the flattened AoS form which would be sent to the GPU, would not be required. However, the latency over-heads of $M$ individual CPU-GPU memcpys (each containing $O(10^2)$ particles) would be overwhelming, especially when considering that a GPU would require $M = O(10^3)$ tasks/computations for efficient execution as demonstrated in Sections 3.2 and 3.4. (Radtke & Weinzierl 2025) have investigated the possible benefits in performance of in-place data structure conversion (AoS to SoA and vice-versa) and (Che et al. 2011) have demonstrated that significant gains could be made from performing such conversions in-situ by the GPU. As such, the data, as-is on SWIFT, requires conversion on the CPU prior to GPU offload as it is not possible to copy the data to the GPU without incurring the aforementioned excessive CPU-GPU communication costs. Additionally, transferring the data as-is would mean that more data is sent than is strictly required for the computation of a single task sub-type (density, gradient, or force) since the particle structs contain the data required for *all* the hydrodynamics computations. Preferably, only the required data for each specific task subtype should be sent to the GPU and only the computation results sent back to the CPU.

To streamline the host-device interaction within the task-parallel approach of SWIFT, we create a new SWIFT task sub-type, which we call "pack_d", "pack_f" and "pack_g" tasks and remove the density, force and gradient tasks as their computations will now be executed by the GPU. These "pack" tasks do not perform computations themselves, but do the following instead:

(i) Copy ("pack") the data needed for the interaction sub-type into CPU side buffers in a layout suitable for GPU execution,
(ii) Trigger the data transfer from CPU to GPU,
(iii) Launch the GPU kernels to perform computations,
(iv) Transfer the results data back to CPU buffers,
(v) Unpack the results from the buffers into the particle data.

We will elaborate on all these steps in what follows.

In the first step, the data for the relevant SPH task from its CPU SoAoS structure is copied into CPU-side data structures in the form of an array of structs (AoS), which is suitable for GPU offload. For each task sub-type (density, force or gradient), each CPU thread constructs an AoS such that each index of the array corresponds to an individual particle struct populated with the particle data required for the computation. Only the data required for specific computation sub-type is copied into this array. As such, there is one array for each task type/sub-type combination to be offloaded. The length of the array corresponds to the number of particles to be offloaded at once. This is dependent on the number of tasks to be offloaded to the GPU concurrently and is a user-set parameter $S_p$, which will be discussed in more detail below. Furthermore, the index of the first and last particles contained in the cells used to perform each SPH computation is recorded. This is necessary to control which particles interact with which particles in the array on the GPU. Entries in this array each correspond to a struct of the form:

```
/* particle data for density */
struct part_send_d{
  float4 x_h; // position vector and h
  float4 v_m; // velocity vector and mass
  int2 cf_cl; // first, last part index in cell
} part_send_d;
```

where `float4` and `int2` are cuda/HIP data types which store four 32-bit `floats` and two `ints` contiguously and allow for vectorisation of load/store instructions to and from the GPU's DRAM to core-local registers. Vectorisation helps saturate the memory bandwidth with efficient transactions maximising bandwidth and computation efficiency. Similar `part_send_f` and `part_send_g` structs are defined to contain data for the force and gradient loops, respectively.

Sending the data one task at a time ($O(10^3)$ bits) on pathways designed for $O(100)$ GB/s throughput is extremely inefficient as this cannot saturate the CPU-GPU communication, introducing a significant overhead and latency of hundreds of thousands of memcpys[2] to the GPU per time step. We therefore implemented an approach where task data is bundled into larger data sets and transferred to the GPU asynchronously. These data sets contain the data for $O(10-10^2)$ tasks, depending on how large the cells are to which tasks are attached and the level in the cell tree at which the optimal SPH computations for these tasks are performed. Generally, the number of particles offloaded on which these tasks act is $O(10^4)$.

**Controlling GPU data transfers and computations via instruction streams**

The `part_send` arrays are designed to contain data for several tasks. Since tasks involve recursive computations, we define a parameter $S_p$ which is the target number of leaf cell interactions we offload to the GPU per cycle. These interactions are herein named leaf computations. If the task-level cell is at a high level in the tree, making recursion a requirement, the task will generate $O(10-1000)$ leaf computations per task, depending on the level in the tree of the original task's cells. Since $S_p$ controls the final size of the data offloaded, it can therefore be pre-defined in a manner to approximately control the number of particles offloaded per cycle. When $S_p$ is larger than the number of leaf computations in a task, this effectively melds leaf computations from several tasks into one large task destined for GPU offload. As such, $S_p$ can be used to optimise data transfers and computations for the available hardware depending on the CPU-GPU interconnect, the GPU memory access speeds, and the number and speed of GPU compute cores.

SWIFT is parallelised using `pthreads`, which means CPU threads are mostly unaware of what other threads are doing and operate independently unless `pthread` synchronisation or communication directives are called. As such, each CPU host thread instructs the GPU independently and requires its own local `part_send` arrays in which to pack the task data. To prevent excessive calls to the computationally expensive process of allocating GPU memory, we allocate each thread's GPU memory once at the start of the simulation. In our approach, we do not require dynamic GPU memory management as the size of data offloaded per offload is fixed to $S_p$ leaf computations. Hence we use CUDA's standard memory allocation procedure (cudaMalloc()). However, advanced memory management techniques are available (see e.g. Wille et al. 2023) and will be investigated in future work for cases where $S_p$ is required to be dynamic to improve offload efficiency. For convenience, we leave the size of the GPU buffer arrays as a user-set runtime parameter, allowing us to easily modify it for various experiment runs. Typically, a size of $O(10^5)$ particles per thread is more than sufficient. With the average size of $\sim$ 44 bytes per GPU buffer part struct (`part_send/part_recv` structs for density, gradient, and force loops), these arrays fit comfortably on GPU memory even for arrays of size $10^6 - 10^7$.

[2] Instructions for data transfer to and from the GPU

**Adapting task bundling and instruction streams to SWIFT's task scheduler**

To maximise concurrency of GPU computations with CPU-GPU communications, many GPU-accelerated codes leverage streaming of instructions wherein each host thread issues groups of instructions to the GPU in different instruction streams. Instructions issued in one stream may be executed concurrently with instructions in another stream. Figure 5 illustrates how this concept may be used to virtually hide the cost in time-to-solution of CPU-GPU data transfers underneath GPU computations.

In each stream, a host to device memcpy ($\mathrm{HD}_n$, where $n$ is the stream ID) is first issued to instruct the GPU to copy the required task data asynchronously with other instructions in different streams. This is followed by instructions to execute the required task computations ($\mathrm{Kernel}_n$) on that data and then by instructions to retrieve the tasks' results data to the CPU via another asynchronous device to host memcpy ($\mathrm{DH}_n$). After the instructions are issued, the GPU is left to perform the memory transactions and compute kernels as and when enough resources are available (GPU copy engines, CPU-GPU link bandwidth and GPU memory and compute cores). If the ratio of computations to memory transactions is sufficiently large, the GPU will be able to hide HD/DH memcpys underneath kernel computations. This can greatly improve off-load efficiency as shown in Figure 5, where a dashed red box highlights a time in the offload process where all CPU-GPU communications are performed concurrently with GPU computation kernels.

To control how and when data is transferred to the GPU, we define a new parameter we call the "bundle size" $S_b$, denoting the number of leaf computations we offload to the GPU via each instructions stream. When we offload $S_p$ leaf computations, we first split them into bundles of $S_b$ leaf computations. Naturally, $S_b \leq S_p$.

Hence, for each GPU offload we launch $S_p/S_b$ kernels and issue HD and DH memcpys for each kernel separately. Therefore, splitting the $S_p$ leaf computations into bundles of size $S_b$ enables the optimisation of the off-load process such that multiple streams are used concurrently to effectively hide CPU-GPU communications.

SWIFT's scheduler controls the creation and assignment of tasks into queues where each CPU/host thread has its own queue, as illustrated in Figure 6. In our off-loading approach, we first add the pack tasks we discussed above to the scheduler such that they are scheduled for execution and placed into queues with other tasks as and when their dependencies are met. If a CPU thread receives a regular (non-GPU-accelerated) SWIFT task, then:

(i) The thread executes the task.
(ii) Upon completion, its data and dependencies are unlocked.
(iii) The thread moves on to execute the next task in its queue.

If, however, a thread receives a pack task, the procedure is slightly more involved and is described as follows:

**A. Dealing with recursive task types**

SWIFT defines tasks on cells higher in the cell hierarchy than would be optimal for SPH interactions based on cells' particle counts. This approach has proven itself performant for temporal and spatial adaptivities necessitated by problems in cosmology. Tasks recursively walk through the large cell's daughter cells to find and compute leaf computations on-the-fly. To minimise task management overheads for the GPU-accelerated code, we implement this recursion as a preliminary step to the data packing process. However, we stress that the recursion only collects a list of leaf computations. It does not pack the particle data. We perform the packing in a subsequent step, outside of the recursion.

The full procedure is outlined as follows:

(i) The host CPU thread receives a pack task from the queue.
(ii) The host thread recurses through the cell associated with the task to identify the list of leaf computations to be executed.
(iii) Once the recursion is complete and a list of leaf computations is created, the host thread begins the packing process, discussed in step B, until $S_p$ leaf computations are packed.
(iv) Once $S_p$ computations are packed, the GPU offload begins following the bundling procedure discussed in step B.
(v) If there are remaining leaf computations in the list that have not been offloaded yet, we pack and offload them following the same procedure repeatedly until we are left with a remainder of leaf computations which is less than $S_p$.

In cases with non-zero remaining leaf computations to offload, the remaining computations are packed into the buffer array, but not offloaded. Instead, the host thread continues to process the next task it obtains from the scheduler's queue. However, the previous task is not marked as completed yet, i.e. its dependencies are not signalled as being satisfied. The next obtained task's data will then be packed together with the previous task's for offloading and the computations performed once the target size $S_p$ is reached. The previous task will finally be marked as completed and its dependencies satisfied once the subsequent GPU offloading cycle containing its data is complete.

**B. Packing and off-load**

(i) The leaf computations have been previously identified through the recursive procedure described in step A. The thread now packs the task's data required for leaf computations into a large array of `part_send` structs. The array is designed to hold $S_p$ computations. We illustrate this in Figure 6, where we use $S_p = 6$ and we set the tasks in the queue at the leaf computation level (operating on cells of width $2H_{max}$) for illustration purposes.
(ii) If the number of leaf computations a thread has packed is smaller than $S_p$, the thread moves on to the next task in its queue. Note that the current task is not marked as completed yet.
(iii) If the thread has packed $S_p$ leaf computations, it triggers the GPU offload process and splits the packed computations into smaller bundles, each containing $S_b$ computations. These bundles are offloaded to the GPU via different instruction streams as and when the GPU's copy engines are available. We illustrate this in Figure 6 using $S_b = 2$.
(iv) The GPU is instructed by the thread to *asynchronously* copy the data for each computation bundle in a separate stream.
(v) The GPU launches compute kernels to perform the computations on each bundle from within the same instruction stream.

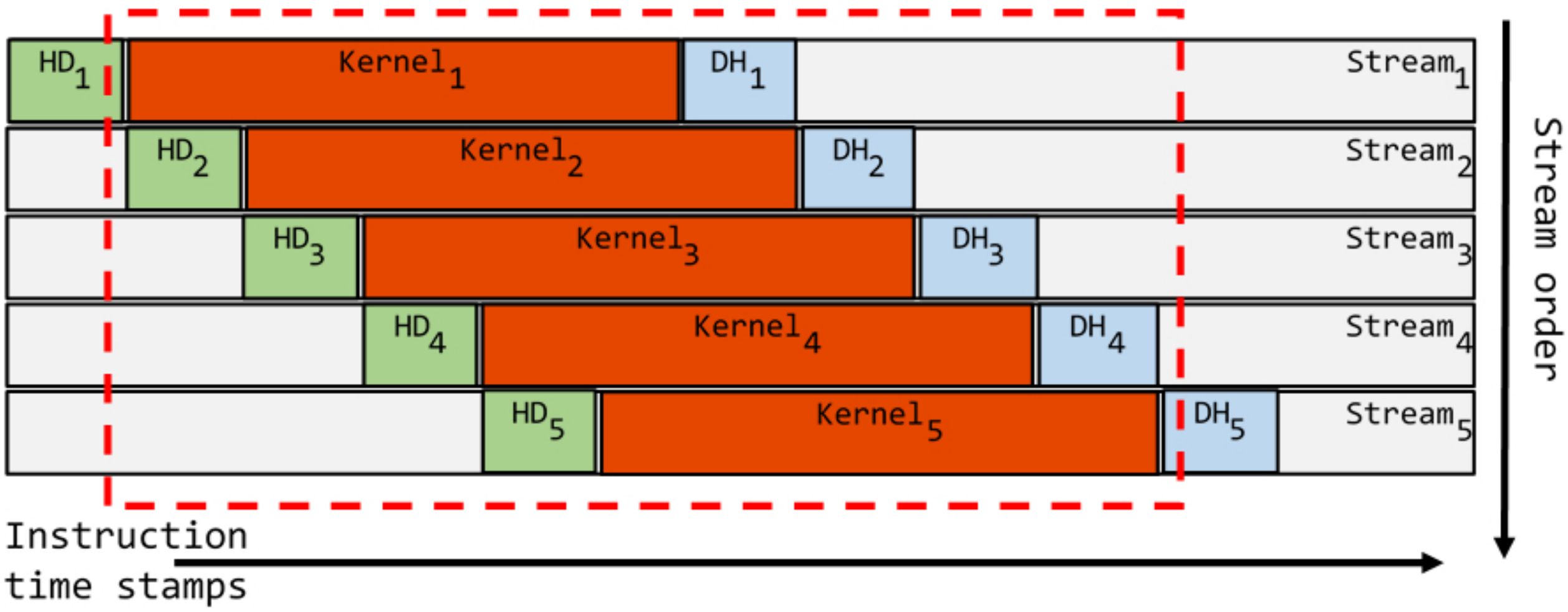

**Figure 5.** Timeline of execution for GPU operations. The schematic illustrates different data transfers and computations executed concurrently by the GPU as issued by one CPU host thread in different instruction streams. HD denotes host-to-device transfer, while DH denotes device-to-host transfer. The host is the CPU node, while the device is the GPU.

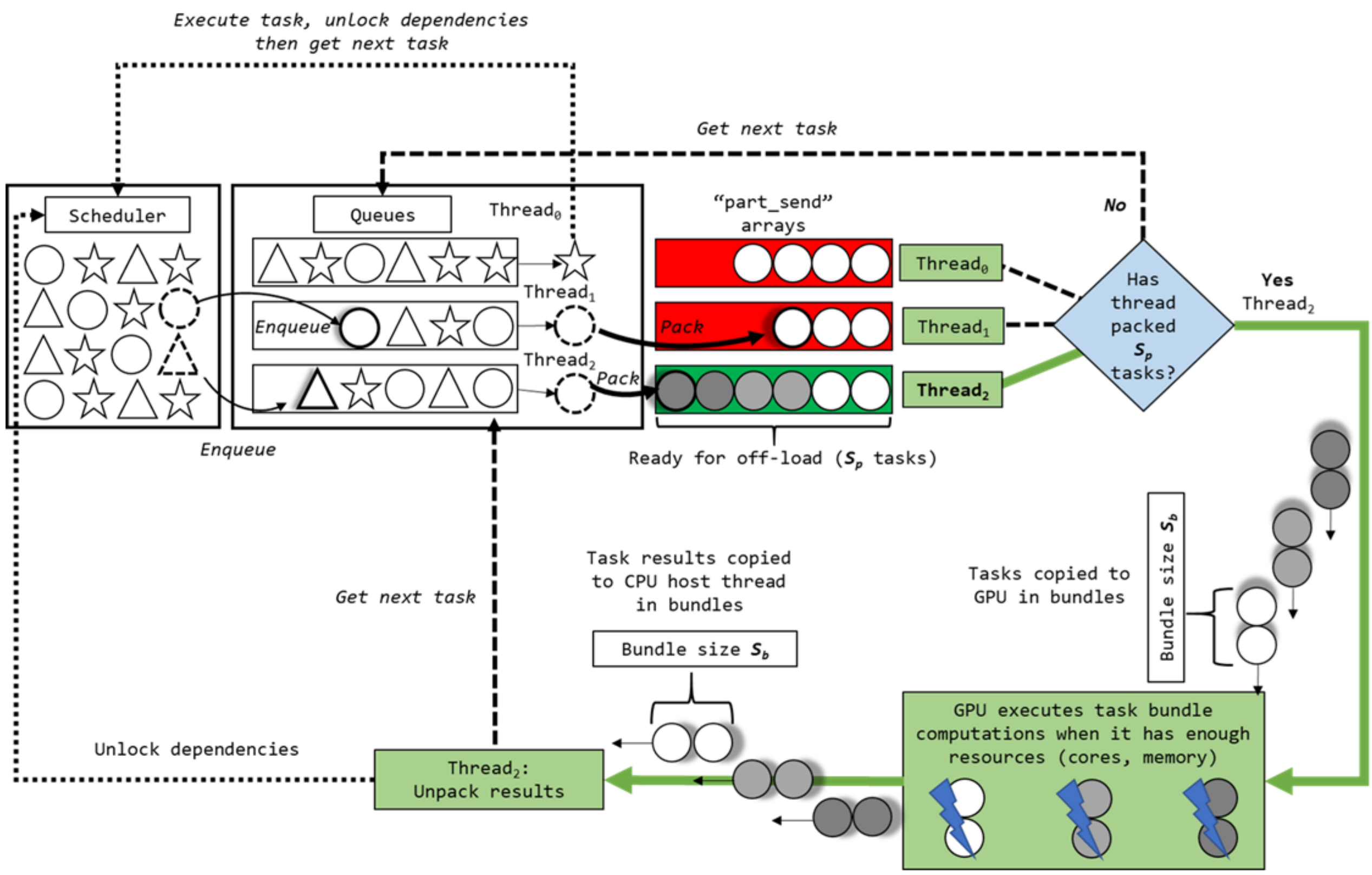

**Figure 6.** GPU-offload mechanism for SWIFT tasks. Circles in this image represent pack tasks where the data required for the computations is packed into arrays of part_send structs. Once a thread has packed the data for $S_p$ leaf computations the offload process begins.

Each instruction stream contains the memcpy instructions of some data and the computation instructions to act on that *same* data (the kernel).

(vi) Asynchronous memcpy instructions are finally issued to copy the results of the computations back to the CPU in the *same streams* from which the computation kernels were launched.

## C. Unpacking

Once the CPU thread receives the results back from the device, it begins unpacking the tasks data: We first loop through the tasks offloaded and then inner-loop over the number of leaf computations belonging to each task. In the DH memcpy and unpacking process, a different data type is used to minimise the size per task of DH communications. It only retrieves the results which the GPU has written to (the computation results) and not the data the GPU read from. Similarly to the `part_send` structs, we define a `part_recv` struct of the form:

```
typedef struct part_recv_d{
  // Density, drho/ dh, w, dw/dh
  float4 rho_dh_wcount;
  // velocity curl vector, divergence.
  float4 rot_vx_div_v;
} part_recv_d;
```

This approach is tested for efficiency in the next Section, where various CPU-GPU types and configurations are used to assess its versatility.

**Data locking**

In SWIFT, when a task is being executed, the thread computing the task has a lock on the cell's particle data it is acting on to ensure no other thread can write to the data to prevent race conditions without relying on atomic operations in performance critical regions. Since the GPU offloading approach proposed here relies on the bundling of multiple tasks, keeping the task data locked after the packing process completes is un-necessary and could introduce inefficiencies by limiting available work. Hence we unlock the task's data once it's been packed and only obtain the lock again once the GPU calculations and the DH memcpy are completed. The resulting data can then be written into the tasks' cells' data in a thread-safe manner.

During this unpacking procedure, the CPU loops through the leaf computation's parent tasks and locks their cell data in a task-by-task basis while writing data to them. This procedure ensures that cells are only locked when necessary, minimising clashes for resource access.

A summary pseudo-code is provided in Appendix A to highlight the steps followed in the proposed GPU-offload method.

## 3 ANALYSIS OF GPU-ACCELERATED TASK COMPUTATION

The proposed GPU-accelerated solver is tested on various GPU computing systems to validate the efficiency of the algorithms across different GPGPU and heterogeneous computing platforms. All results in this Section 3 are conducted using a single GPU.

Three different systems were used for the analyses presented in this section as summarised in Table 1.

### 3.1 Numerical Setup

For all results presented in this paper the test case simulated was the 3-D Gresho-Chan vortex (Gresho & Chan 1990) SWIFT benchmark test case[3] using the SPHENIX scheme (Borrow et al. 2022). In this test, a cubic domain with periodic boundaries is discretised into approximately uniformly spaced particles in a glass-like distribution (generated by allowing a uniform grid of particles to settle to a state where the kinetic energy has stabilised to near-zero) with uniform mass. The domain size is $L^3$ with non-dimensional side length $L = 1$. The smoothing kernel support radius is set to $1.9\Delta x$ where $\Delta x$ is the initial particle spacing to ensure approximately 64 neighbouring particles for each particle. The cubic spline is used for all tests presented in this paper. For all tests presented in Section 3, the domain is split such that each cell contains roughly 64 particles. Therefore the task recursion mechanism is not used. This ensures that each GPU offload only contains either self *or* pair tasks and allows assessment of their efficiency separately. All the density, gradient and force, self and pair tasks are executed on the GPU. The particles are assigned initial velocities according to the analytical solution and the density is set to $\rho_0 = 1$ with polytropic index $\gamma = 5/3$. The axis (centre) of the vortex is parallel to the $z$-axis (the analytical velocity is zero in the $z$-direction) and the radius of the vortex $r$ is parallel to the $x$-$y$ plane.

The analytical azimuthal velocity ($v_f$) and pressure ($p$) only vary in the $r$ direction. The analytical solution is

$$v_f(r) = v_0 \begin{cases} u & u \le 1 \\ 2-u & 1 < u \le 2 \\ 0 & u > 2 \end{cases}, \tag{27}$$

$$p(r) = p_0 + \begin{cases} \frac{1}{2} v_0^2 u^2 & u \le 1 \\ 4v_0^2 \left( \frac{u^2}{8} - u + \ln u + 1 \right) & 1 < u \le 2 \\ 4v_0^2 (\ln 2 - \frac{1}{2}) & u > 2 \end{cases} \tag{28}$$

where $u = r/R$ with the constants $R = 0.2$, $v_0 = 1$ and $p_0 = 0.5$ in our setup. Further details of this test can be found in Rosswog (2015).

### 3.2 Acceleration on Nvidia V100 GPU with IBM POWER9 CPU

The N8 CIR system (Bede[4]) on which we conduct the tests in this subsection has very similar architecture to Summit[5] with V100 GPUs connected to IBM POWER9 CPUs via a dedicated 50GB/s NVSwitch port which is over 3 times faster than the V100's 16 channel PCIe 3 port (16 GB/s). For the V100 tests, 3 different resolutions are used ($N_p = 64^3$, $128^3$ and $256^3$ particles) to assess the effect of problem size on GPU off-load efficiency as compared to CPU-only baselines using a full POWER9 CPU with 32 cores. In the tests presented here, $S_p$ and $S_b$ are varied for each resolution to determine the optimum combination of off-load parameters which ensure maximum concurrency of GPU computations with CPU-GPU data transfer. Figure 7 shows the results for the velocity and pressure fields in the Gresho-Chan vortex test confirming correctness of the GPU implementation (results obtained from the original CPU only SWIFT with the same setup are identical).

To assess the effects of total offload size and offload size per bundle on the efficiency of the GPU computations, a battery of tests is performed where the total number of the computations offloaded $S_p$ and the total number of computations per bundle $S_b$ are varied. Increasing $S_p$ has the effect of minimising the average cost of kernel launch overheads (per leaf computation) as we launch fewer kernels with increased $S_p$. On the other hand, having more bundles per offload (reducing $S_b$) has the effect of increasing kernel launch overheads (as we launch more kernels) but allows for additional concurrency of computations and CPU-GPU communications as discussed in Section 2.3 and illustrated in Figure 5 and Figure 6. Therefore, we investigate the optimal combination for $S_p$ and $S_b$ which minimises computation time overall. Moreover, as different GPUs have different capabilities, the ideal combination of $S_p$ and $S_b$ may vary between GPU models, further motivating the tests presented here. The range of $S_p$ values to be tested is specified and $S_b$ is varied from $S_b = S_p/8$ to $S_b = S_p$. This process is currently a partially manual endeavour where a bash script is set up to launch a batch of tests which execute

[3] The case is available online gitlab.cosma.dur.ac.uk/swift/swiftsim

[4] https://n8cir.org.uk/bede

[5] www.olcf.ornl.gov/summit/

**Table 1.** Summary of systems used for the single GPU acceleration tests.

| System | CPUs | GPUs | CPU-GPU Link |
|---|---|---|---|
| Bede | 32 nodes with 2x IBM POWER9 32 core CPUs each and 512GB DDR4 RAM | Nvidia V100 with 32GB memory. 4GPUs per node. | NVLink 2.0 (50 GB/s) |
| | 5 nodes with one 72 core Nvidia Grace CPU each and 480 GB LPDDR5X RAM | Nvidia H100 with 96GB memory. 1 GPU per node. | NVLink C2C (450 GB/s) |
| Gemini 2 | 1 node with two 28 core Intel Xeon Gold 6330 CPUs and 128GB DDR4 RAM | Nvidia L40 with 48GB memory. 4 GPUs per node. | PCIe4 (32GB/s) |

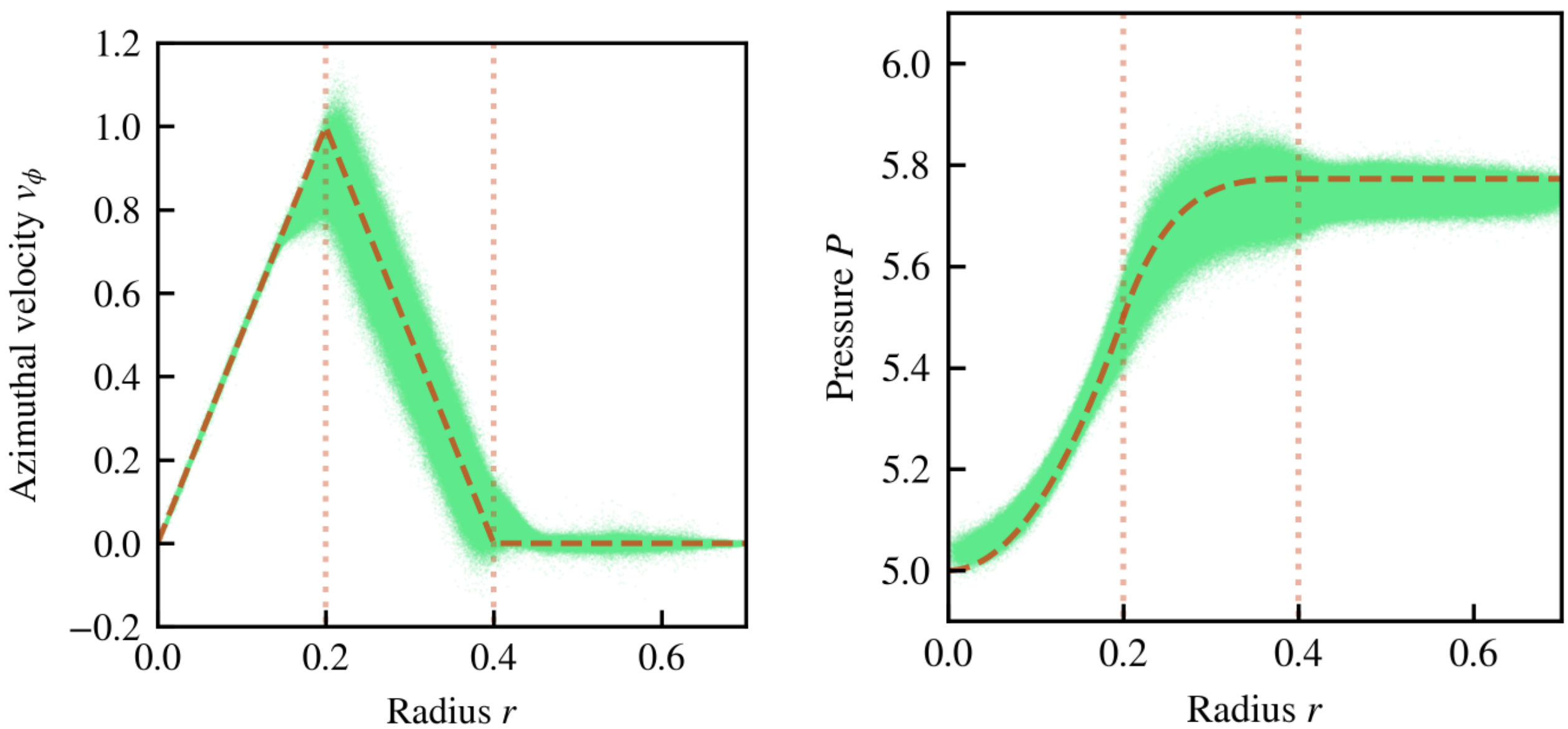

(a) Velocity profile

(b) Pressure profile

**Figure 7.** 3-D Gresho-Chan vortex test results using $256^3$ particles. Velocity (a) and pressure (b) profiles are plotted against vortex radius. Dashed profiles indicate the analytical solution and green dots indicate the numerical solution predicted by SWIFT for all particles using their radial position $r$ with respect to the vortex centre at $t = 0.1$.

10 time steps and report the average compute times for self and pair tasks. However, we aim to automate this process for the first release of GPU-accelerated SWIFT by developing a stand-alone benchmark which would report the optimal configuration for the hardware it is executed on. The total number of SPH particles allocated to compute nodes remains approximately constant in SWIFT simulations due to adaptive domain partitioning and load balancing. Therefore, the optimisation process is only required at the start of the simulation.

Figure 8 shows the compute time results for the lowest resolution test with $64^3$ particles where increasing the total amount of data off-loaded to the GPU $S_p$ improves the efficiency of the off-load and thereby reduces the overall data transfer and compute time. For the best performing case ($S_p = 64, 128$ for the self tasks and $S_b = 32, 64$ for pair tasks) the total off-load time is 0.221 s and 0.0105 s for the pair and self tasks, respectively, as compared to 0.263 s and 0.0539 s when computed on the CPU.

Increasing the resolution to $N_p = 128^3$ improves the GPU efficiency as shown in Figure 9 where the best efficiency is achieved with $S_p = 256, 512$ and $S_b = 128, 256$ for the pair and self tasks, respectively. The CPU times for $N_p = 128^3$ are 1.756 s and 0.318 s as compared to the most efficient GPU off-loads with 0.971 s and 0.0477 s for the pair and self tasks, respectively.

Increasing the resolution to $N_p = 256^3$ improves efficiency further as seen in Figure 10. For this test, 32 host threads were used for the GPU time results. The best performing cases are for $S_p = 1024, 2048$ and $S_b = 256, 512$ for the pair and self tasks, respectively, where the CPU times were 12.7 s and 2.14 s whilst the GPU times were 4.07 s and 0.197 s for the pair and self tasks, respectively.

The best achieved speedups ($T_{\rm CPU}/T_{\rm GPU}$ where $T_{\rm CPU}$ is the total time to compute the tasks on the CPU and $T_{\rm GPU}$ if the total time to pack the tasks, compute them on the GPU and unpack the tasks) are summarised in Figure 11 which shows speedups for the 3 resolutions compared to the time for 32 CPU threads to perform the computations for consistency across resolutions.

To expand on the GPU performance for the sets of V100 tests, Figure 12 shows GPU activity profiles obtained using Nvidia's NSight Systems[6] profiler for the cases with $N_p = 64^3$, $128^3$ and $256^3$. They illustrate how the GPU executes instructions such as memcpy and computation kernels as instructed by the host. Select time windows in the computations were chosen to illustrate cases (combinations of $N_p$, $S_p$ and $S_b$) where concurrency of CPU-GPU communications was achieved (thereby hiding communication costs) and cases where concurrency was not achieved.

[6] `developer.nvidia.com/nsight-systems`

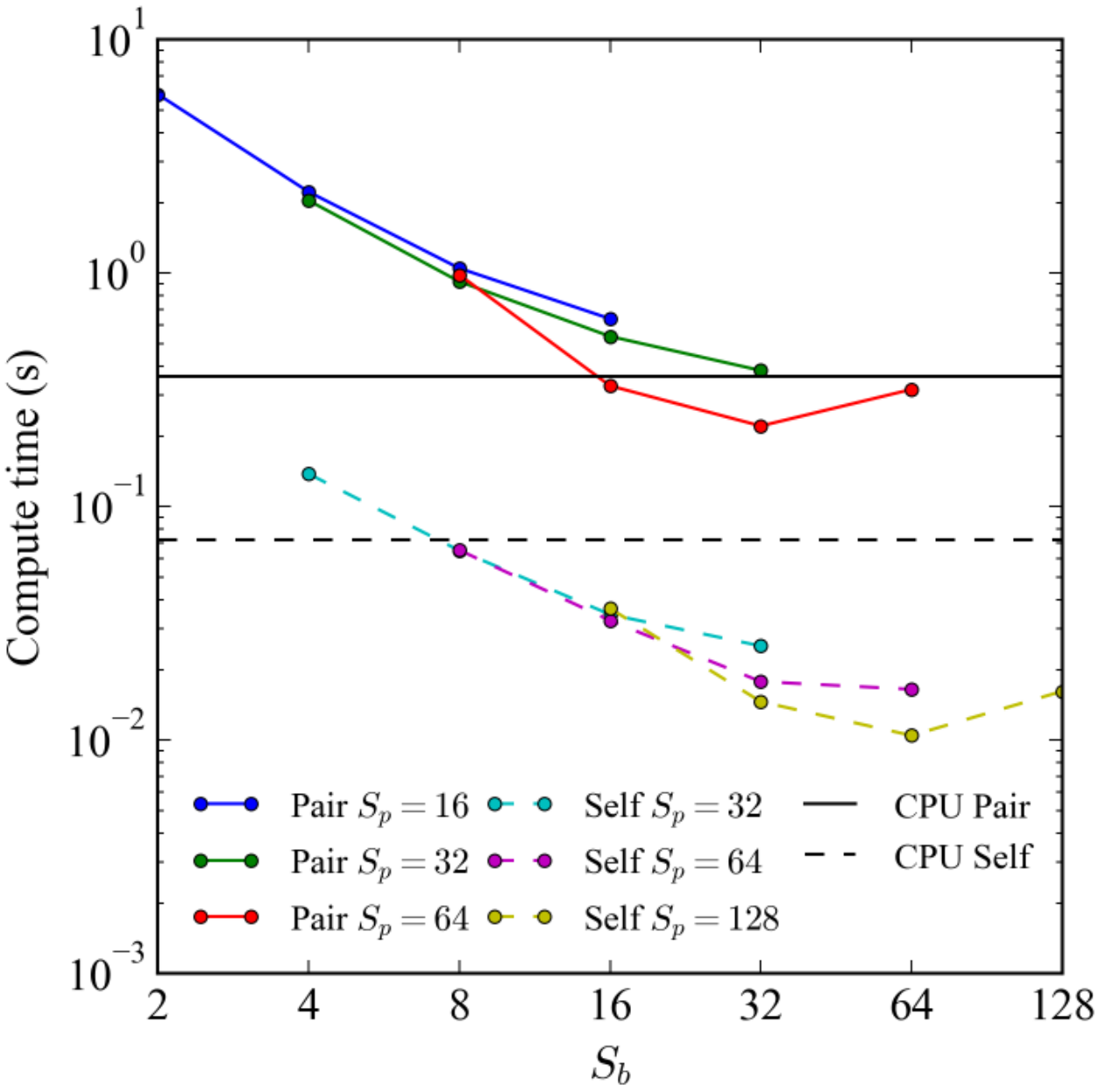

**Figure 8.** Variation of compute time on V100 with $S_p$ and $S_b$ for the case with $N_p = 64^3$ using 16 host threads compared to CPU times with 32 threads.

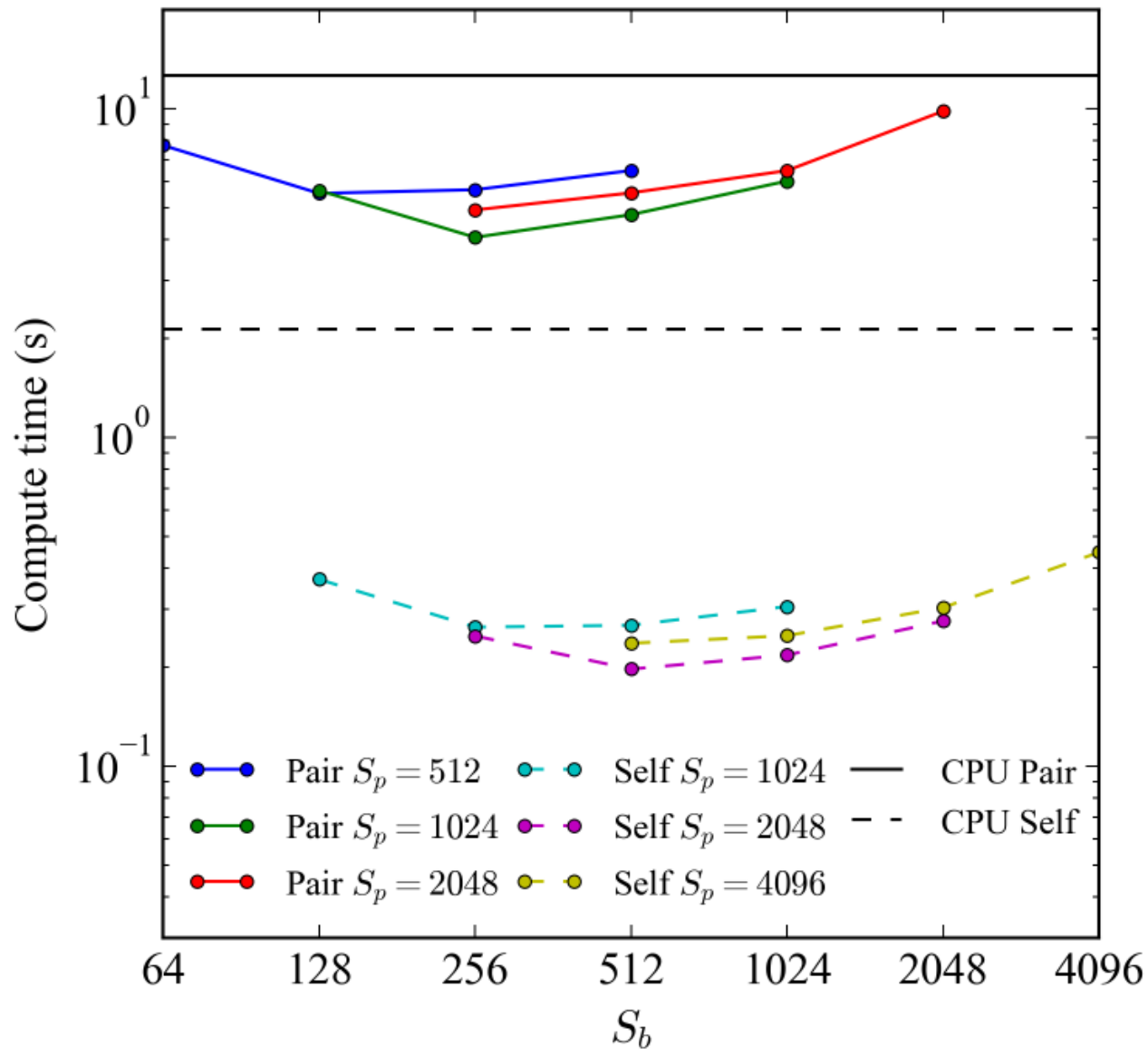

**Figure 10.** Variation of compute time on V100 with $S_p$ and $S_b$ for the case with $N_p = 256^3$ using 32 host threads compared to CPU times using a full POWER9 CPU with 32 threads.

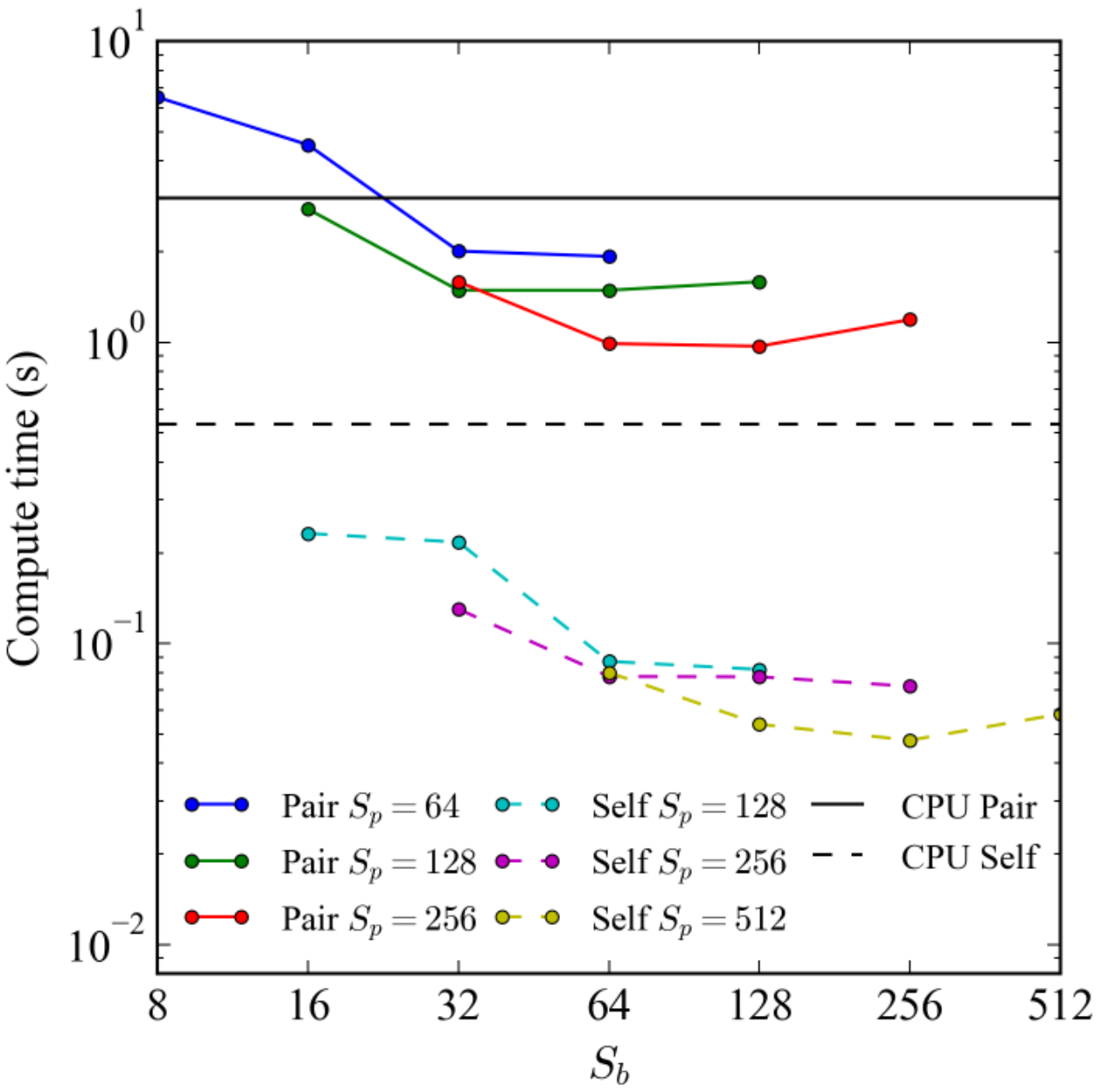

**Figure 9.** Variation of compute time on V100 with $S_p$ and $S_b$ for the case with $N_p = 128^3$ using 16 host threads compared to CPU times with 32 threads.

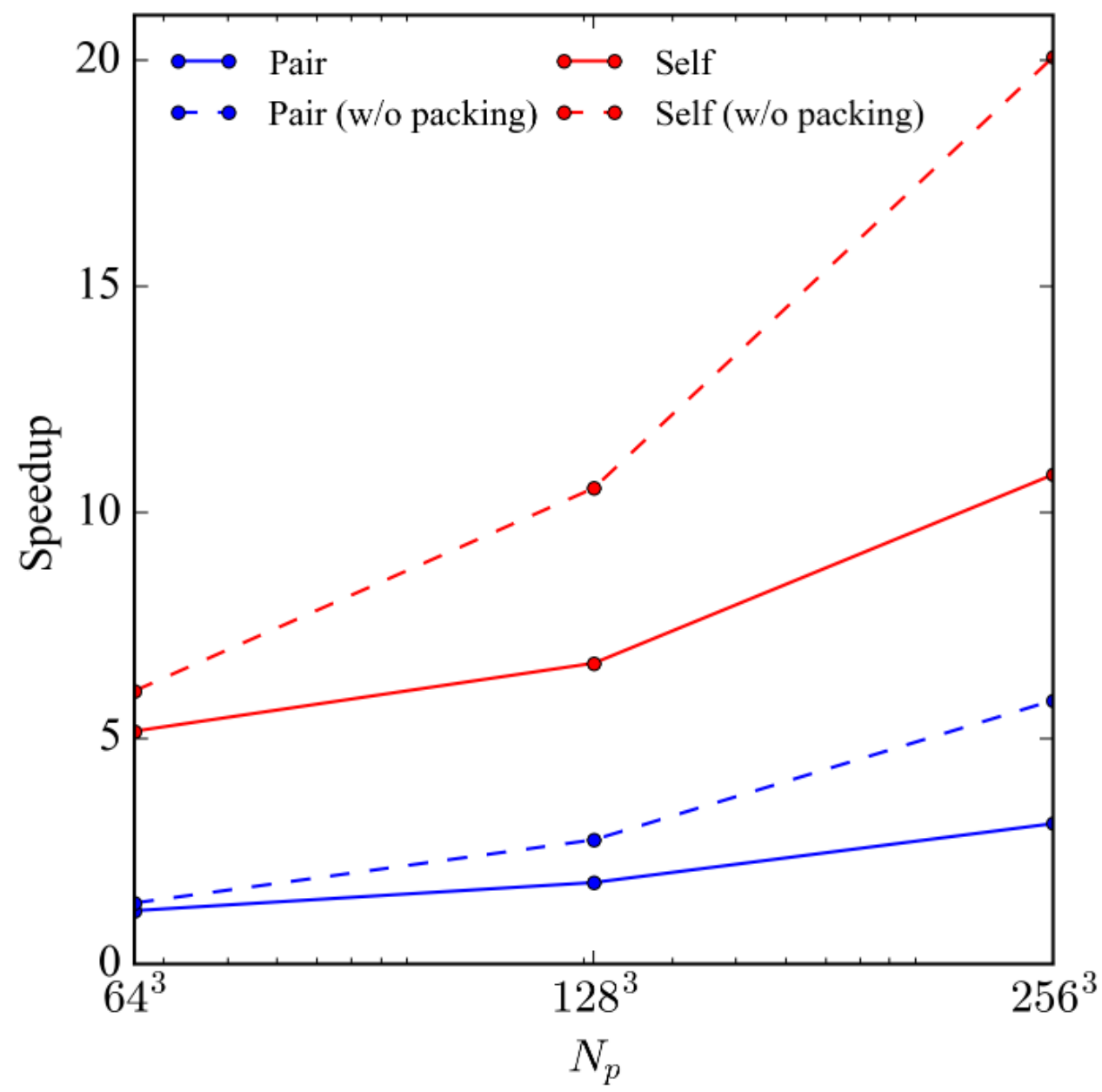

**Figure 11.** The best speedups achieved for self and pair tasks executed on a V100 GPU for three different values of $N_p$. For each resolution the best speedup is plotted as compared to 32 POWER9 CPU threads.

Figure 12a and 12b show that for the cases with $N_p = 64^3$ and $128^3$ concurrency is achieved: memcpys are mostly hidden underneath compute kernels. However, only four kernels are executed per cycle, i.e. two bundles of tasks from two host threads are executed per offload cycle, which limits the ability of the GPU to efficiently schedule the execution of kernels and memcpys concurrently. The major limitation here is that there are not enough tasks available to saturate the GPU-offload mechanism. Saturation requires that enough CPU threads have sufficient tasks to pack and provide a continuous stream of sizeable data flowing to/from the GPU whilst the GPU is busy performing computations.

On the other hand, Figure 13 shows that for the best performing case with $N_p = 256^3$, data transfer and kernel execution are highly concurrent with two-way data transfers occurring concurrently and concurrent execution of kernels and memcpys. For the self task sample offload cycle in Figure 13a, there are eight kernels executed meaning that (four bundles of tasks from two host threads) which are executed concurrently. For the pair tasks in Figure 13b, there are

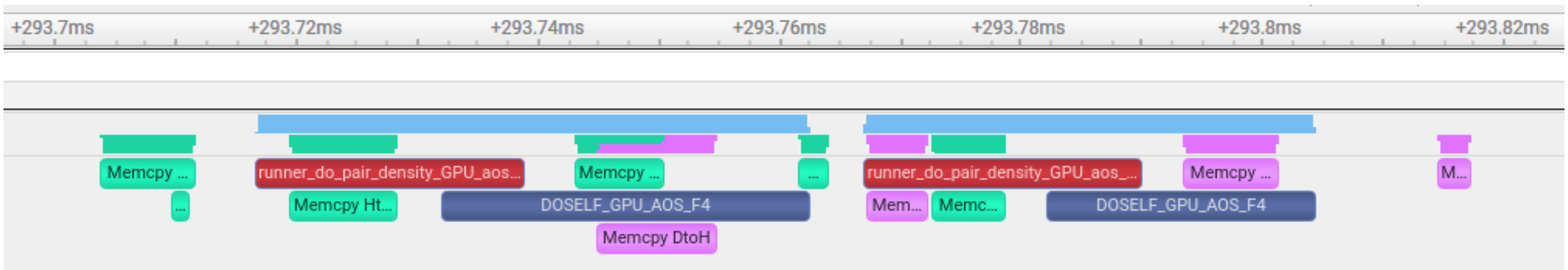

(a) $N_p = 64^3$, $S_p = 128$ and $S_b = 64$

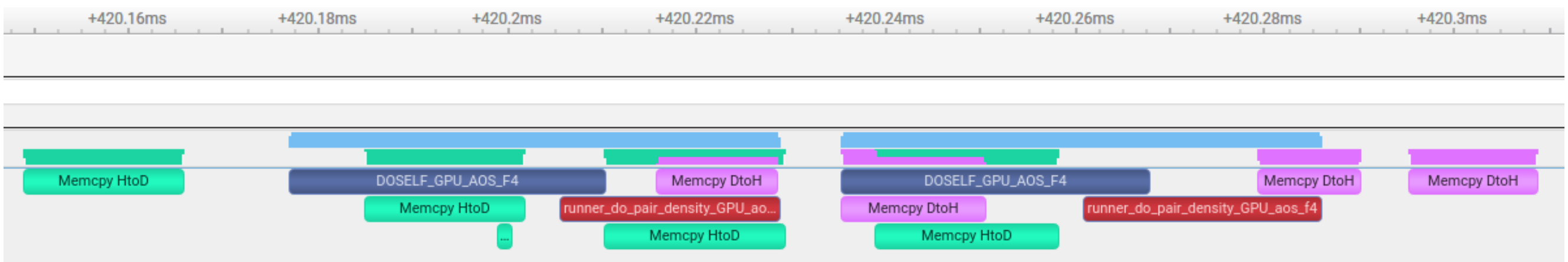

(b) $N_p = 128^3$, $S_p = 512$

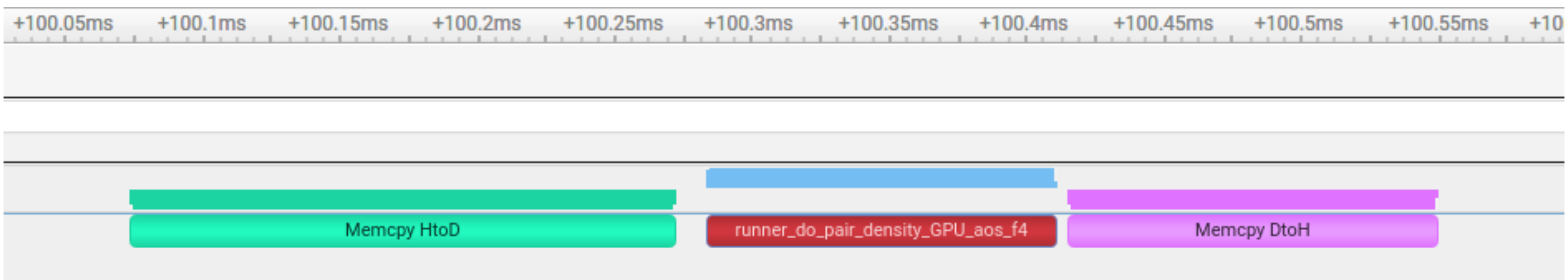

(c) $N_p = 64^3$, $S_p = 128$ and $S_b = 64$

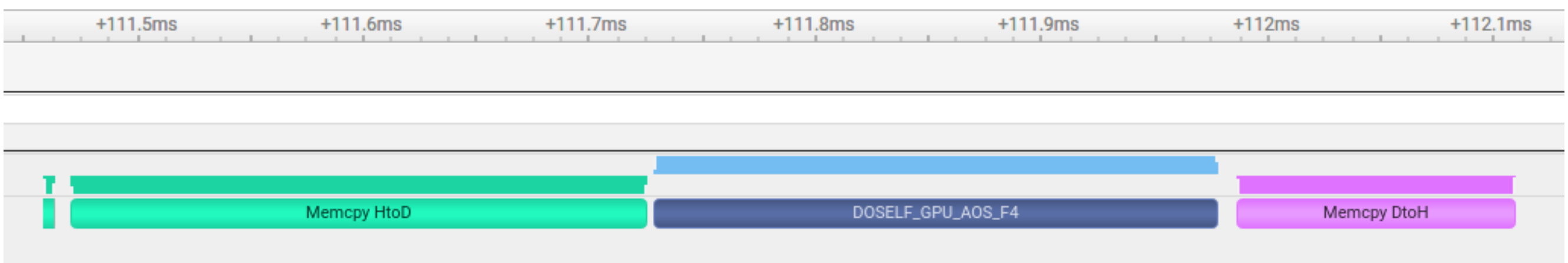

(d) $N_p = 128^3$, $S_p = 512$

**Figure 12.** Profiles for a GPU offload cycle on the V100 GPU for select test cases. (a) $N_p = 64^3$, $S_p = 128$ and $S_b = 64$ with 16 host threads. (b) $N_p = 128^3$, $S_p = 512$ and $S_b = 256$ with 16 host threads. (d) and (c) $N_p = 256^3$, $S_p = 4096$ and $S_b = 4096$ with 32 host threads. Figure (c) shows a profile of pair tasks and (d) shows a profile of self tasks. Green blocks represent host to device data transfer, lavender blocks represent device to host data transfer, red blocks represent kernels executing pair tasks and dark blue blocks represent kernels executing self tasks.

**Table 2.** Average times required to complete self and pair tasks and maximum data transfer (memcpy) throughput for selected test cases with varying $N_p$. The columns for $N_p = 64^3$ and $128^3$ show the best performing tests for those resolutions whilst the two columns with $N_p = 256^3$ show the best and worst performing combination of $S_p$ and $S_b$ for that resolution *from Figure 10.*

| | $\mathbf{N_p = 64^3}$ (Self $S_p = 128$, $S_b = 64$) (Pair $S_p = 64$, $S_b = 32$) | $\mathbf{N_p = 128^3}$ (Self $S_p = 512$, $S_b = 256$) (Pair $S_p = 256$, $S_b = 128$) | $\mathbf{N_p = 256^3}$**, best** (Self $S_p = 2048$, $S_b = 512$) (Pair $S_p = 1024$, $S_b = 256$) | $\mathbf{N_p = 256^3}$**, worst** (Self $S_p = S_b = 4096$) (Pair $S_p = S_b = 2048$) |
|---|---|---|---|---|
| Self kernel time/task | 0.237 $\mu s$ | 0.0654 $\mu s$ | 0.0641 $\mu s$ | 0.0664 $\mu s$ |
| Pair kernel time/task | 0.285 $\mu s$ | 0.0816 $\mu s$ | 0.0859 $\mu s$ | 0.0293 $\mu s$ |
| Memcpy throughput | 20 GB/s | 45 GB/s | 47 GB/s | 56 GB/s |

sixteen kernels (tasks from four host threads) being executed in the same cycle which allows for even greater concurrency.

The GPU throughputs for computations and CPU-GPU communications for all tests presented in this sub-section are highlighted in Table 2. Figure 12c and Figure 12d show the most extreme examples presented here. For the case with $N_p = 256^3$ and $S_p = 4096$ and $_b = 4096$, memcpys and kernels are the most efficient as shown in Table 2. However, the overall offload time (see Figure 10) is the longest of all tests conducted using $N_p = 256^3$ (0.448s and 9.84s for self and pair tasks respectively which is not significantly faster

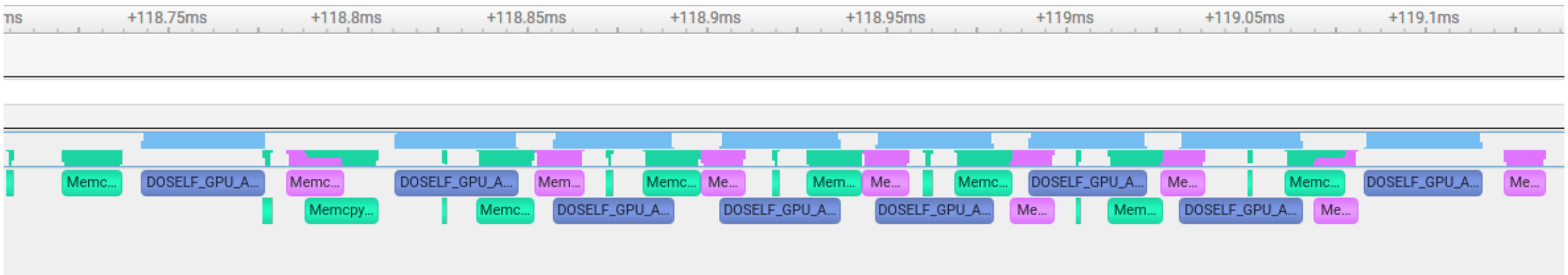

(a) Profile for a GPU offload cycle for self tasks

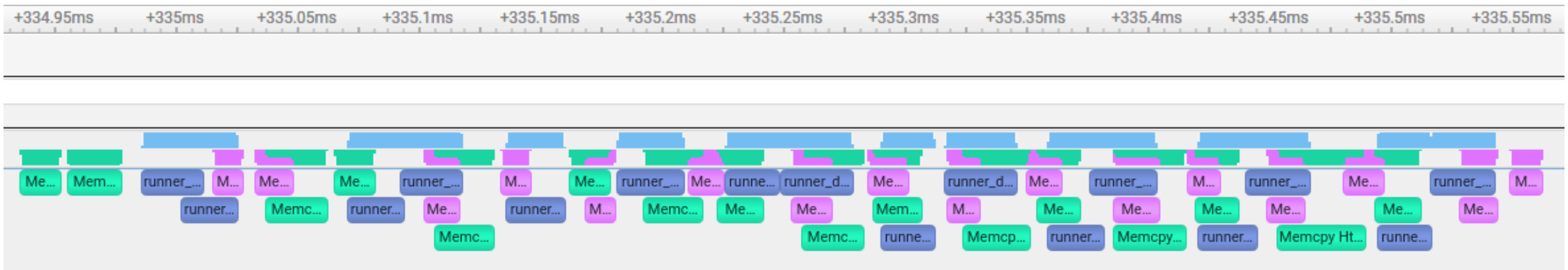

(b) Profile for a GPU offload cycle for pair tasks

**Figure 13.** Profile for a GPU offload cycle on the V100 GPU for the case with $N_p = 256^3$, $S_p = 2048$ and $S_b = 512$. (a) Self tasks. (b) Pair tasks. Green blocks represent host to device data transfer, lavender blocks represent device to host data transfer and dark blue blocks represent kernels executing self or pair tasks.

than the CPU time of 2.14 s and 12.7 s for self and pair tasks on 32 CPU threads). The reason why this case is slower overall than the best performing case ($N_p = 256^3$ , $S_p = 2048$, $S_b = 512$) is due to the fact that no concurrency between CPU-GPU communications and GPU computations was achieved.

Further analysis presented in Table 2 illustrates that the time required to compute a leaf computation by the GPU improves as $N_p$, $S_p$ and $S_b$ are increased. The same holds for the memcpy throughput measured by NSight Systems which exceeds the stated 50 GiB/s throughput of the NVlink2 connection for the largest pack and bundle sizes tested. This is due to the larger amount of computations and communications saturating the GPU's capabilities in terms of CPU-GPU communications bandwidth and also due to the fact that kernel launch overheads are minimised. The number of GPU instructions is reduced, and the total size of the data transferred is increased, by a factor of 4. However, since no concurrency of data transfer and computations is achieved, the overall off-load efficiency is significantly deteriorated.

Figure 11 clearly demonstrates that the packing and unpacking of data on the host is taking a considerable amount of time especially for the largest test with $N_p = 256^3$ particles where it is roughly 47% of $T_{\rm GPU}$. An issue with the packing and unpacking is that the compiler is unable to vectorise the data access due to the non-contiguous nature of the data on the host. Vectorisation of the host memory should speedup the packing and unpacking process for single precision (32 bit) data on CPUs with 512-bit Advanced Vector Extensions (AVX-512). In theory, 16 data copies can be performed concurrently on such a configuration provided the data is contiguous in memory. However, data bus latency and bandwidth would likely limit the achieved speedups to less than 16 times. Investigating what speedups are possible if the packing/unpacking process is vectorised is of significant interest. However, this requires an overhaul of the host code data structure, for the CPU solver and for other SWIFT task types which use the SPH task data and is the focus of future work (Ivkovic et al., In prep.).

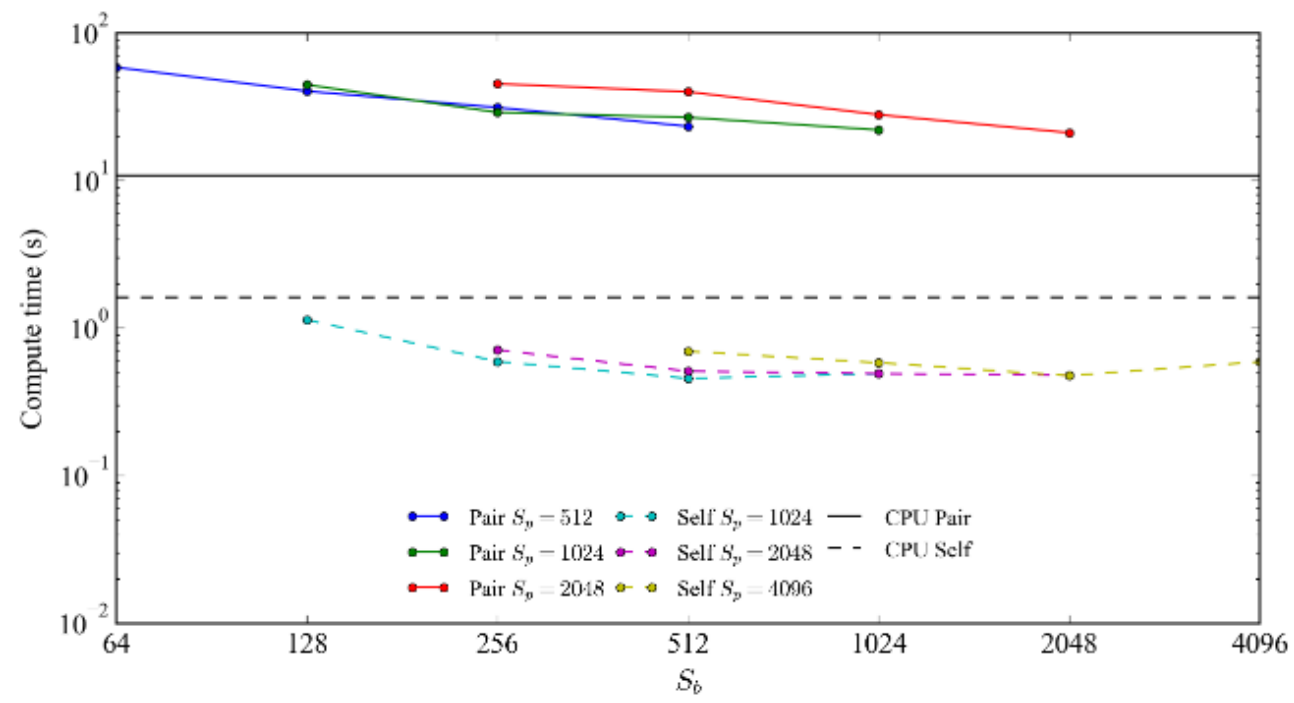

**Figure 14.** Variation of compute time on L40 GPU with $S_p$ and $S_b$ for $N_p = 256^3$ using 32 host threads compared to CPU times with 32 threads.

### 3.3 Acceleration on Architectures With Conventional PCIe Connections

In this section, the GPU-accelerated SWIFT solver is tested on the University of Manchester compute system (Gemini 2) with Nvidia's L40 GPUs, where connectivity to CPUs is via 16 PCIe 4 channels which can deliver 32 GB/s CPU-GPU data transfer speeds. The setup for these tests is identical to the setup for the tests in Section 3.2 with $N_p = 256^3$.

For the L40 GPU runs the performance was significantly affected by the speed of the CPU-GPU connection. Figure 14 shows that whilst the self tasks were still faster on the L40 GPU, in comparison to 32 Intel Xeon Gold 6330 CPU threads, the pair tasks were in general slower on the GPU in comparison; the best achieved time for pair task completion ($S_p = 2048$ and $S_b = 2048$) was 21.2 s in comparison to 10.8 s. For the self tasks the best achieved time was 0.458 s in comparison to 1.62 s on 32 CPU threads. This is due to the fact that the GPU is unable to hide the CPU-GPU data transfers under computations due to the much slower PCIe connection being unable to keep up with the computational throughput of the L40 GPU. Since for the tests in this Section the pair tasks require twice the amount of data as for the self tasks and the number of particle interactions is generally much smaller, especially for pair tasks acting on two cells connected by a corner, for example, the CPU-GPU data transfer speed becomes the dominant bottleneck.

Figure 15 shows profiling results for the density tasks illustrating that data transfer and compute concurrency is minimal. This issue is usually alleviated by keeping the SPH particle data on the GPU for the duration of the simulation but with task parallel solvers such as SWIFT this is not possible since the tasks are randomly assigned to each thread by the scheduler in order to minimise thread idle time on the CPU: a cell's density tasks may be assigned to one thread whilst it's force and gradient tasks may be assigned to another thread or stolen from the thread's queue by another thread.

Figure 15 shows that although concurrency is achieved between data transfers to and from GPU memory for these tests, concurrency with computations is minimal due to their relatively short duration. Hence, the GPU-accelerated solver performs poorly on this machine.

### 3.4 Acceleration on Nvidia Grace-Hopper superchip

The Grace-Hopper superchip nodes on the N8CIR Bede cluster are used to perform this set of tests. The Grace-Hopper (GH200) consists of Grace (a 72 core ARM CPU) connected to Hopper (an H100 GPU) via specialised NVLink fabric capable of 450 GB/s (in each direction).

For the GH200 tests $N_p = 256^3$ particles and $S_p$ was varied from 512 to 2048 for the pair tasks and 1024 to 4096 for the self tasks. $S_b$ was also varied with four values used for each $S_p$ to assess GPU offload efficiency. Figure 16 shows the compute times when tasks were executed on the H100 GPU with 32 host threads as compared to task execution on 32 threads of the Grace CPU. 32 threads were used to ensure consistency and comparability with all other analyses presented in Section 3.2.

The Grace CPU is roughly 2.75 and 2.54 times faster in executing the pair and self tasks, respectively, than the POWER9 CPU whilst the best performing case executed on the H100 GPU for pair and self tasks is 2.05 and 2.06 times, respectively, faster than the V100 (Figure 7). The maximum speedups achieved when computing tasks on the H100 vs 32 threads of the Grace CPU are roughly 2.3 and 8.8 which are similar to the V100 tests as compared to the POWER9 CPU. When not including the packing and unpacking times the H100 GPU speedups are 6.6 and 23.8 when compared to 32 threads on the Grace CPU. The patterns followed by the GPU offload efficiency with variation of $S_p$ and $S_b$ are very similar to the patterns observed with the V100-POWER9 combination discussed in Section 3.2. This is expected as the architecture is similar due to the use of NVLink instead of PCIe to connect the CPU to the GPU. Analysing the overall time required to offload to the GPU for the best performing test with $S_p = 4096$ and $S_b = 1024$ and $S_p = 2048$ and $S_b = 512$ for the self and pair task types, respectively, the ratio of packing and unpacking time in comparison to the total offload time was approximately 65% further highlighting the importance of optimising the packing and unpacking process.

The SWIFT CPU solver test was repeated to use the full Grace CPU with 72 threads in order to compare the full Grace CPU with the Hopper GPU. When using the full CPU (72 threads), the self and pair tasks were completed in 0.374 s and 2.09 s, respectively. In comparison to the time required to complete the tasks using 32 threads (0.838 and 4.61 s) this is a near linear speedup, 2.24 and 2.21 times faster. The speedups for the GPU-accelerated code executed on a Hopper GPU in comparison to execution a full Grace CPU using 72 cores are 1.04 and 3.93 for the pair and self tasks, respectively. Whilst the speedups for the self tasks are promising, the pair tasks are not accelerated by GPU execution and this is due to the pair tasks requiring larger data transfers to the GPU whilst performing less computations than the self tasks. Since the CPU already has the data to hand and uses the highly efficient pseudo-Verlet list (Gonnet et al. 2013) discussed in Section 2.2.2 for the pair tasks, which are not implemented in the GPU computations, the current implementation is unable to provide a significant speedup for the pair tasks. Also, these tests were conducted without task recursion enabled and the tasks were attached to leaf-level cells (of size $\sim 2H$). This was to isolate and highlight the difference between the efficiency of self and pair leaf computations on the GPU but does not reflect the full solver's performance with recursion enabled which is discussed in Section 4.

### 3.5 Summary of Analyses

The results of the analyses in Section 3 are summarised in Table 3 where the best performing cases are presented. Table 3 shows that the H100 and V100 GPUs connected via proprietary connections perform significantly better than the L40 GPU connected to the CPU via PCIe. Table 3 also shows that for the best performing GPUs the packing/unpacking process is a significant portion of the GPU offload process.

## 4 SOLVER PERFORMANCE FOR A FULL SIMULATION

For the analyses presented in Section 3, the focus was on assessing the performance and potential of GPU acceleration of SWIFT's SPH tasks. In this Section, a higher level analysis is presented focussing on overall performance on the Grace-Hopper superchip for a full simulation where non-SPH task types are executed on the CPU and where the overheads of task management are also assessed.

In the GPU-accelerated tests presented in this Section, the recursion mechanism discussed in Section 2.2.2 is enabled for the pair tasks. However, for the self tasks recursion is not enabled. Recursion for GPU-offloaded self tasks has been implemented but it is currently not performative due to excessive packing and unpacking overheads on the CPU associated with a current bottleneck arising from duplicate data being packed and transferred to/from the GPU. This bottleneck is discussed in Section 4.5 where paths to eliminating it are also discussed. Herein, self tasks are therefore treated as leaf computations; packed and offloaded as they are when a CPU thread receives them. As such, self leaf computations for the tests presented here contain $\sim 64$ times more particles than the ideal size for SPH computations.

### 4.1 SWIFT CPU Code Performance

The Gresho-Chan Vortex simulation, presented in Section 3, is used with $N_p = 256^3$ particles for the analyses presented here.

The unmodified SWIFT CPU code provides a base-line for comparison with the GPU-accelerated code. Figure 17 shows the analysis for a representative time step, illustrating when each task is executed by which CPU thread. For this case, the initial conditions are set such that the domain is decomposed into $16^3$ top-level cells. The task recursion method as discussed in Section 2.2.2 is used to perform the SPH computations such that each particle in a task only searches for neighbours within the ideal cell size which for this case would contain ~64 particles.

At the start of the step, there is a small gap of roughly 39 ms. This is the section where CPU threads are processing book-keeping

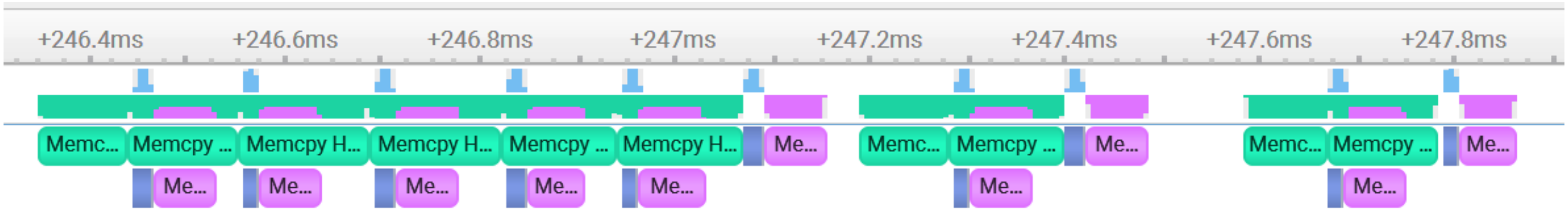

(a) Pair tasks with $S_p = 1024$ and $S_b = 512$

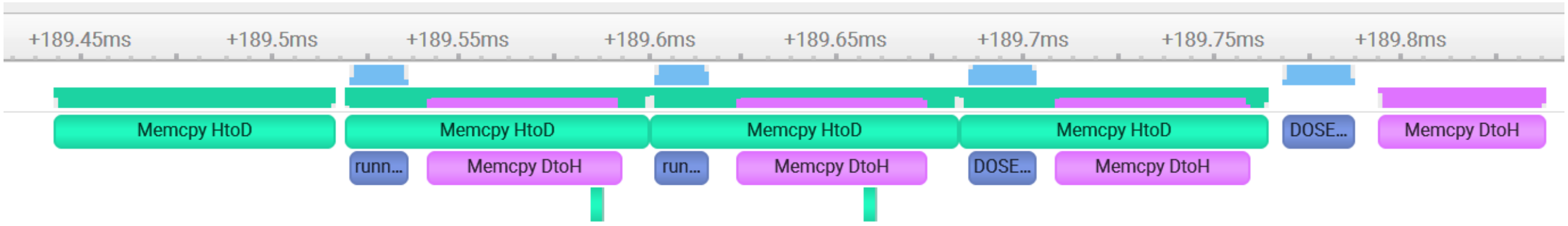

(b) Combination of self and pair tasks scheduled for execution on the GPU by two different host threads.

**Figure 15.** (a) Pair tasks with $S_p = 1024$ and $S_b = 512$. Memcpys to the GPU and CPU are coloured green and pink respectively. Computation kernels are coloured blue. (b) Combination of self and pair tasks scheduled for execution on the GPU by two different host threads.

**Table 3.** Hardware performance summary. All CPU results are using 32 threads for consistency. The speedups presented are based on the time required to compute all the SPH task types (self and pair) and subtypes (density, gradient and force) offloaded to the GPU.

| Hardware | Total time (self and pair tasks) | Packing and Unpacking time | Best combination of $S_p$ and $S_b$ | Speedup with packing | Speedup without packing |
|---|---|---|---|---|---|
| 2 POWER9 CPU (32 cores) | 14.8 s | N/A | N/A | 1 | 1 |
| Nvidia V100 GPU | 4.27 s | 1.99 s | $S_p = 2048$, $S_b = 512$ | 3.47 | 6.49 |
| Grace CPU (72 cores) | 5.45 s | N/A | N/A | 1 | 1 |
| Nvidia H100 GPU | 2.07 s | 1.34 s | $S_p = 4096$, $S_b = 1024$ | 2.63 | 7.47 |
| 2 Intel Xeon Gold 6330 CPU (28 cores) | 12.5 s | N/A | N/A | 1 | 1 |
| Nvidia L40 GPU | 23.8 s | 5.17 s | $S_p = 1024$, $S_b = 512$ | 0.52 | 0.67 |

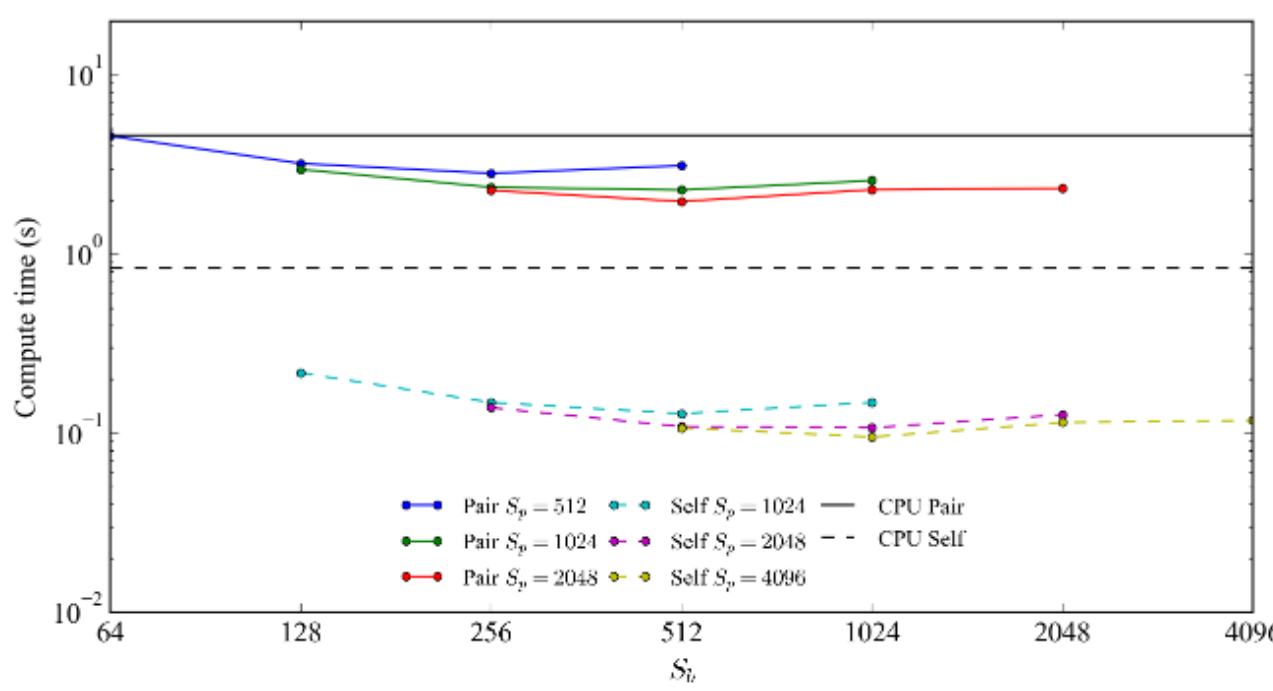

**Figure 16.** Variation of compute time on GH200 with $S_p$ and $S_b$ for $N_p = 256^3$ using 32 host threads compared to CPU times with 32 threads.

activities on tasks required for the time step. The full Grace CPU completes the time step in 2032 ms. The sum of each thread's wall clock time required to complete all the computationally intense SPH tasks (density, force and gradient task subtypes) was recorded as 137026 ms, translating to 1903 ms per thread to estimate the time-to-solution for the SPH tasks. The initial task management time is hence ~ 2% of the total time step. The number of particles updated through a full time step per second per CPU core may be used to provide a reference for comparison with other solvers. For the cases outlined above, this was ~115 thousand particles marched through one time step per second per core (8.26 million particles updated by the full CPU per second).

### 4.2 GPU-Accelerated Code Performance

The test in Section 4.1 is repeated using the GPU-accelerated solver with 72 host CPU threads and $S_p = 2048$ and $S_b = 512$ for pair task types whilst for the self task types $S_p = 64$ and $S_b = 16$ . This ensures roughly the same number of particles are offloaded for self and pair tasks since the self leaf computations contain 64 times more particles than the pair leaf computations. The task execution for a representative time step is presented in Figure 18. Figure 18 shows that the initial task management time is 30 ms on average (2.7% of total time).

The total time required to complete the time step with the GPU-accelerated code was 1132 ms and the time required to complete all the SPH tasks was recorded as 66876 ms (929 ms when averaged over the number of host threads of 72). The packing/unpacking time was ~ 80% of this time which further emphasises the need to eliminate the data duplication bottleneck discussed in Section 4.5. Converting the time-to-solution to particle updates per second gives ~14.8 million particle time step updates per second for this test on a single Grace-Hopper superchip. Considering that the GPU-offload mechanism used here is still under development (see Section 4.5 for details of upcoming developments) 14.8 million particle updates per second per superchip is an excellent starting point, especially when compared to state-of-the-art SPH GPU solvers where the particle data resides permanently on the GPU such as SHAMROCK (David–Cléris et al. 2025) which achieves 20 million updates per second on an Nvidia H100 GPU[7] (the same GPU used in the Grace-Hopper superchip). We stress that for the computations presented here the full CPU of the Grace-Hopper chip is also used for the packing/unpacking process and for other task types which are not GPU-accelerated.

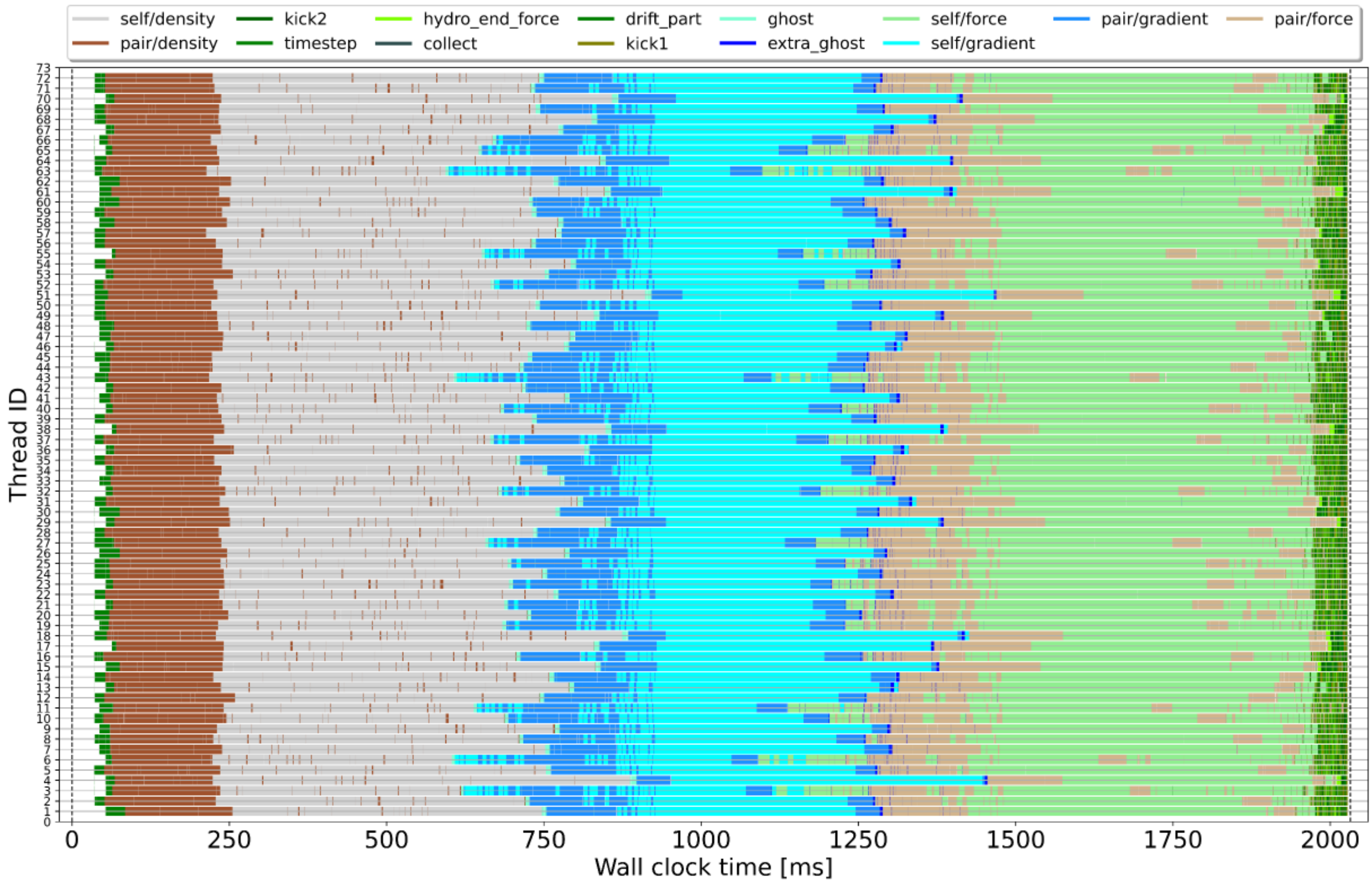

**Figure 17.** Tasks performed in a time step by the unadulterated SWIFT CPU code on the Grace ARM CPU using 72 threads. Each row illustrates which task type each CPU thread is executing at that point in time.

There is notable idle time after the force (gpu_pack_f) sub-types as shown in Figure 18 towards the end of the step. The total time where threads were not executing computational tasks was 5110 ms (roughly 71 ms when averaged over 72 threads) as compared to the full Grace CPU results (3764 ms or 52 ms when averaged over 72 threads). It is currently unclear exactly what causes this idle time after the force tasks (gpu_pack_f), but it is clear that some threads are unable to retrieve tasks from queues. Since leaf computations are offloaded to the GPU in packs of size $S_p$, the tasks dependent on these $S_p$ leaf computations must wait until the results are copied back onto the CPU memory before they are unlocked and then enqueued, as illustrated in Figure 6. This leaves some queues empty as they wait for tasks to be unlocked, while some tasks may be in the queue but prevented from running due to data access conflicts, as discussed in Section 2.2. The GPU performance for task execution (929 ms on average) is still an improvement (a speedup of 2.05) in comparison to the thread averaged time required for the Grace CPU to complete the SPH tasks with 72 threads using the original highly optimised SWIFT solver, 1903 ms. Comparing the total time required to complete the entire time step (including inter-task idle time), the GPU code achieves a speedup of 1.8. The results of the analyses in Sections 4.2 and 4.1 are summarised in Table 4.

### 4.3 Multi-GPU Code Performance

To investigate the performance of the GPU-accelerated solver on multiple CPUs and GPUs, the same Grecho-Chan vortex test is conducted on the HPC advisory council's Thea cluster's Grace-Hopper system, which is composed of six Grace-Hopper superchips. At the time of writing, only four of its "superchips" were online. The resolution is set to $512^3$ particles (8 times higher than in the previous test) and the case was simulated for ten steps four times using 1 to 4 full Grace-Hopper chips (residing on distinct nodes) to validate the solver's performance on multiple GPUs by recording the average time required to complete a time step. As such, for each test, $72 \times N_{GPUs}$ CPU cores were used to host $N_{GPUs}$. We assign one MPI rank per Grace-Hopper with inter-node communications passing through the system's InfiniBand network. Speedup is measured as the average time required to complete 1 time step compared to the time required to complete 1 time step using a single Grace-Hopper superchip. (Figure 19) shows that the solver achieves near-perfect strong scaling for this test case. The speedups were 2.1, 2.8 and 4 for the tests with 2, 3 and 4 Grace-Hoppers, respectively. These tests demonstrate that the proposed GPU-offloading mechanism does not hinder SWIFT's scalability for small scale tests. However, testing on $O(100-1000)$ GPUs is still required to definitively validate the GPU-accelerated solver's scalability. At the time of writing, the authors were unable to extend the scalability testing beyond the tests presented here as access to large scale machines was unavailable. This will be subject of future work.

[7] https://shamrock-code.github.io/

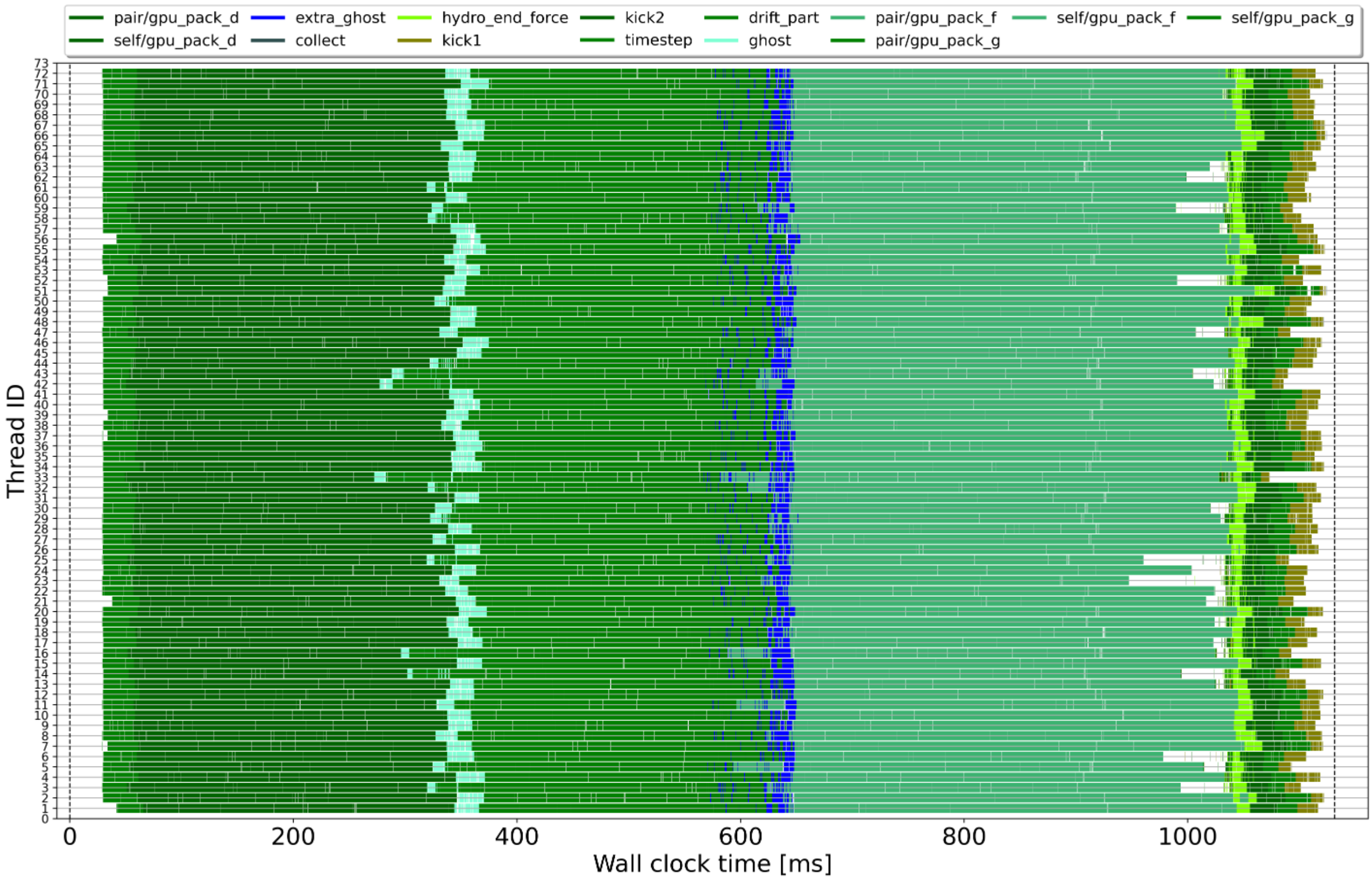

**Figure 18.** Tasks performed in a time step by the GPU-accelerated SWIFT code. Each row illustrates which task type each CPU thread is executing at that point in time. Tasks labelled gpu_pack are the pack tasks for the density task subtype, where the data required for the GPU computations is packed into part_send arrays before offloading to the GPU once $S_p$ leaf computations are packed. gpu_pack_g and gpu_pack_f are the pack tasks for the gradient and force task subtypes, respectively. The dashed vertical line at the right hand side of the Figure is the end of the time step.

**Table 4.** Summary of CPU and GPU code performance for a full simulation.

| Hardware | Total step time | Initial task management time | Total time spent outside task execution | Time spent executing all density, gradient and force task subtypes | Number of particles updated through a time step per second |
|---|---|---|---|---|---|
| Nvidia Grace CPU (72 threads) | 2.03 s | 39 ms | 52 ms | 1.9 s | 8.26 million |
| Nvidia H100 GPU | 1.13 s | 30 ms | 140 ms | 0.93 s | 14.8 million |

### 4.4 Energy consumption

For highly parallel applications, GPU-acceleration could lead to significant improvements in energy efficiency (Qasaimeh et al. 2019). To confirm whether GPU-acceleration improves the energy efficiency of SWIFT's SPH solver, the Gresho-Chan test is repeated using a number of resolutions ($128^3$, $256^3$, $512^3$ and $6^3 \cdot 128^3$). Ten time steps are simulated and SLURM is used to monitor the energy consumed by each test. The odd resolution of $6^3 \cdot 128^3$ was used because the CPU memory available on the Grace-Hopper superchip (480GB) could not fit $1024^3$ ($\approx$ 1 billion) particles. Note that the GPU memory used for the largest test with ~450 million particles was < 20 GB (21% of the 96GB available on the Grace-Hopper's H100 GPU). For this set of tests, we use one Grace-Hopper superchip. For each resolution, we measure the energy consumed when running SWIFT only on the Grace CPU with 72 CPU threads and compare this to the energy consumed when using the full superchip.

As shown in Figure 20 (a) - where energy usage is plotted in log scale for clarity - for the smallest test with $128^3$ particles, using only the Grace CPU is significantly more energy efficient, consuming 1.09 kJ in comparison to 6.31 kJ consumed when using the full superchip. This is due to the overheads of offloading to the GPU outweighing parallelisation efficiency gains since the problem is too small to efficiently utilise the GPU. However, when we increase the resolution to $256^3$ the energy consumed by the full superchip is slightly lower (by ~ 4%) than the energy consumed when running the solver only on the CPU (12.3 kJ vs 11.8 kJ). Further increasing the res-olution to $512^3$ shows that the energy savings increase to $\sim$ 29% (105.1 kJ used by the CPU vs 75.06 kJ used by the full superchip) as shown more clearly in Figure 20 (b) where energy usage is plotted in linear scale. The energy saved remains consistent at $\sim$ 29% for the largest case tested with $6^3 \cdot 128^3$ particles, where the CPU required 442.21 kJ whilst the simulation executed by the full superchip required 312.25 kJ. The tests presented here demonstrate that, provided enough computations are offloaded to the GPU to ensure it is operating efficiently, significant energy savings are possible with our proposed GPU-acceleration mechanism.

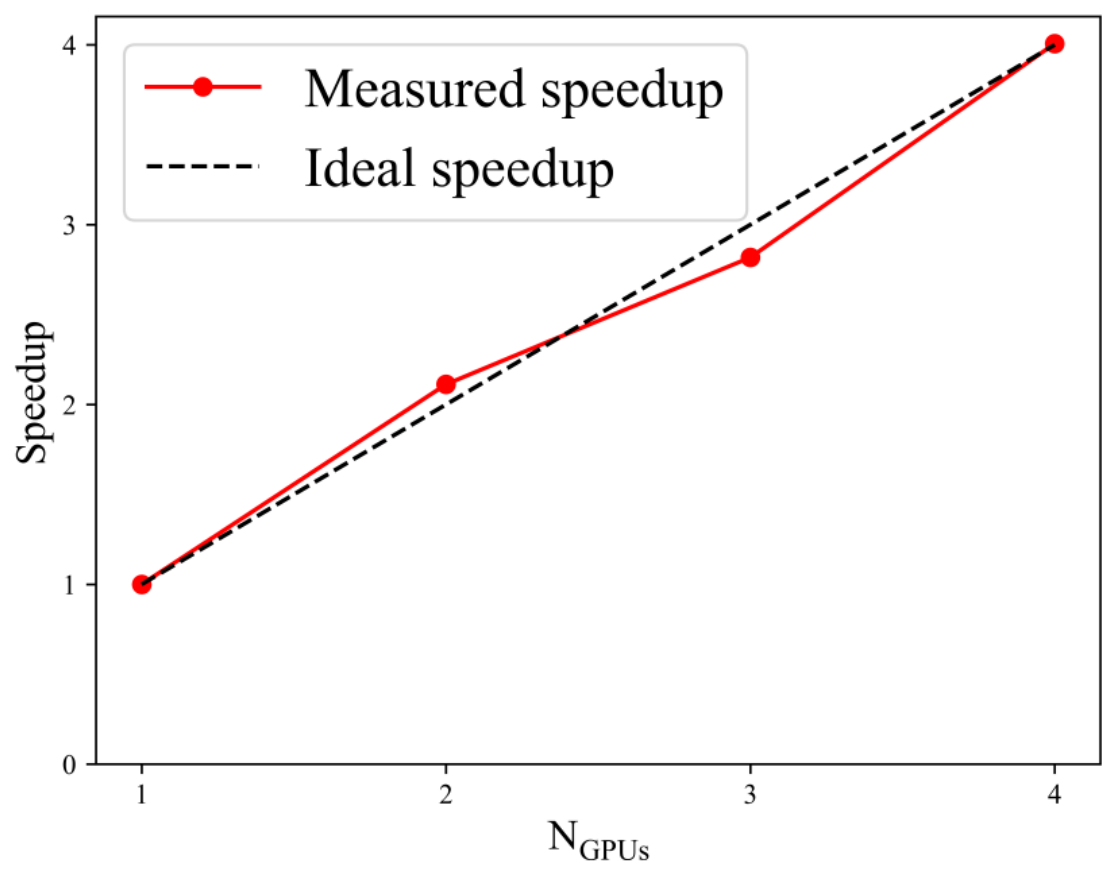

**Figure 19.** GPU-accelerated SWIFT scalability on Grace-Hopper superchips with $512^3$ particles. The baseline is the average time required to complete 1 time step using 1 superchip and 72 CPU cores to host the simulation.

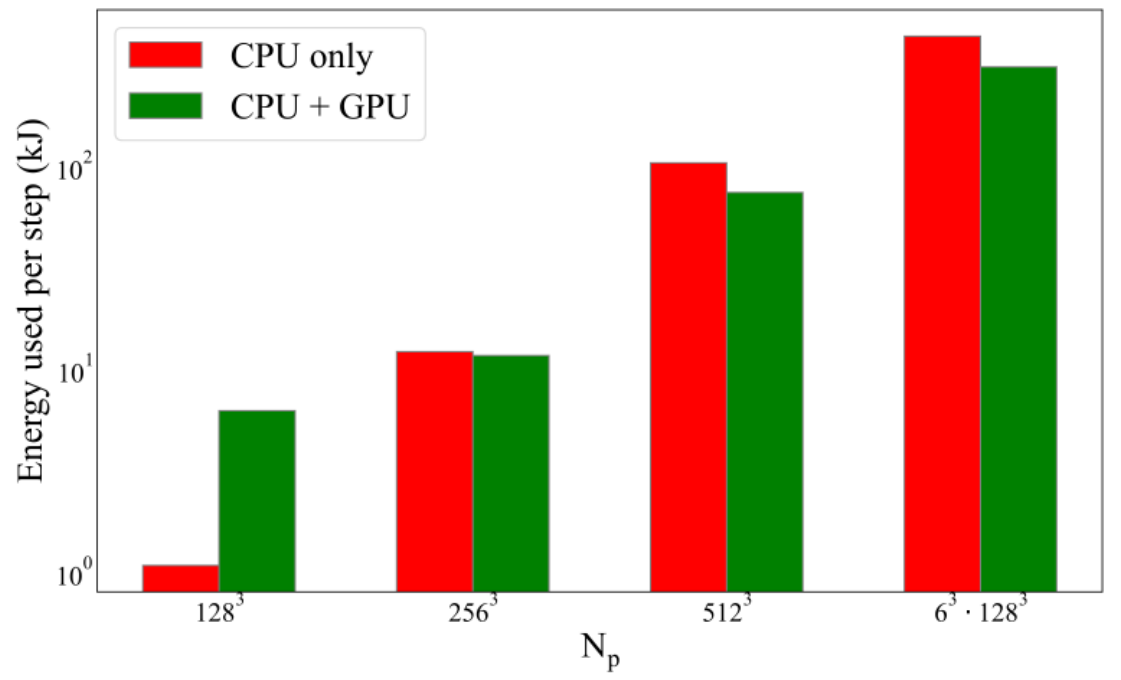

(a) Log scale for consumed energy.

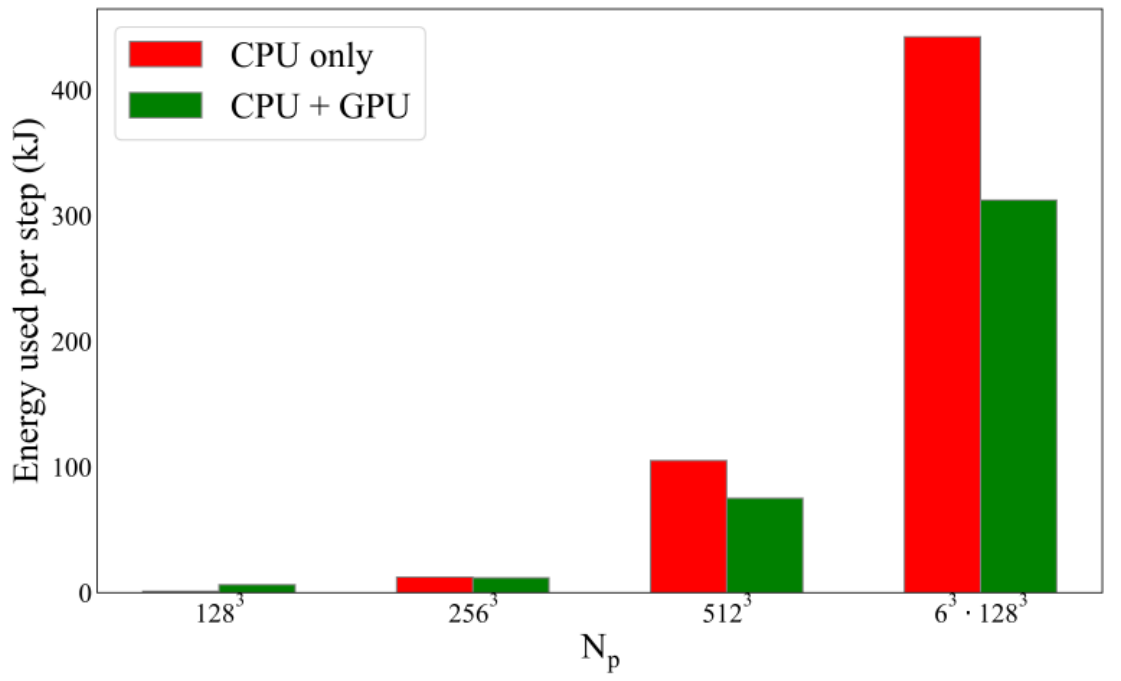

(b) Linear scale for consumed energy.

**Figure 20.** Energy consumed by the full Grace CPU (CPU) vs the full Grace-Hopper superchip (CPU + GPU) to complete ten time steps for tests with different resolutions plotted with log (a) and linear (b) scales for energy consumed per step. GPU-accelerated SWIFT scalability on Grace-Hopper superchips with $512^3$ particles. The baseline is the average time required to complete 1 time step using 1 superchip and 72 CPU cores to host the simulation.

### 4.5 Addressing Performance Limitations

In the current SWIFT SPH solver, the data structures used to access and store the particle data are too complex to be efficiently copied across to the GPU without re-organising them into simpler data structures first. This reorganisation, which we called "packing" and "unpacking", has been a major factor limiting the achievable speedups. As discussed in Section 3.3, when the CPU-GPU data transfers and the GPU computation kernels are at their most efficient, the packing and unpacking process becomes a major bottleneck, taking up to 65% of the total time required to prepare and offload the computations to the GPU. The ratio goes up to $\sim$ 80% for the full simulation tests in Section 4.2. We will address this bottleneck in future work through multiple avenues.

Firstly, we will eliminate duplicate data packing and unpacking during a single offload cycle. In our current implementation, each computed leaf cell interaction requires its own instance of cell/particle data to be present on GPU memory, adding a significant overhead for the packing and unpacking process as well as for the CPU-GPU communications. This data duplication arises from the fact that a task's leaf cells need to undergo several interactions. To illustrate, consider the leaf cell (red dashed square) in the bottom left corner of Figure 4. Within the depicted task, this leaf cell will undergo pair-type interactions with its 3 adjacent leaf cells. Hence, in our current implementation, its data will be packed, copied, and unpacked 3 times. While this approach currently allows us to avoid race conditions (as each GPU thread works on performing the neighbour interactions for a single particle), it clearly contributes severely to our most pressing performance bottleneck. Indeed, we measured that on average only 10% of the leaf cells in the data set for each GPU offload were unique.

Whilst the elimination of this data duplication is anticipated to increase GPU computation time - synchronised reduction algorithms for example will be required to avoid race conditions due to multiple GPU threads writing to the same data - we expect that an efficient implementation will ensure that the savings gained in packing/unpacking times and CPU-GPU communications will significantly outweigh the increase in GPU computation time.

Secondly, we will investigate various CPU-side memory layouts for particle data to optimise the buffering procedures needed by CPU-GPU data transfers. We expect that we can gain additional significant performance in the packing and unpacking procedures by splitting the CPU-side particle struct into smaller structs optimised for the GPU data transfer (Ivkovic et al., in prep.).

And thirdly, should the two approaches above prove insufficient, we will furthermore investigate retaining on the GPU (subsets of) the particle data already transferred to it until the end of a simulation step. This can be achieved for example by allocating globally shared arrays on GPU memory (accessible by all CPU threads) to store the data received by the GPU, as opposed to the CPU-thread-local buffers we use now. On the CPU side, a cell whose data has been transferred during a simulation step already can hold information in a flag signalling that subsequent packing and CPU-GPU transfer of the same data can be skipped. Since we would still be transferring only subsets of particle data which are strictly necessary for the computation of the hydrodynamics interactions on the GPU (84 bytes per particle instead of 160 for SPHENIX SPH), this approach would still leverage the larger CPU memory to fit bigger problems on a single node, albeit less so than with our current approach.

## 5 CONCLUSIONS AND OUTLOOK

In this paper, the first steps towards a novel framework for GPU-acceleration within a task-parallel SPH solver with next-gen "super-chips" have been established. The main goal of this work was to develop an efficient offloading mechanism for particle hydrodynamics embedded within SWIFT's task-parallel engine. While SWIFT's fine-grained tasks are optimised for CPU execution, their small size is sub-optimal for GPU execution. Hence, at its core, our approach is to bundle these fine-grained tasks into larger packets on-the-fly as and when they are made available by the task scheduler, improving the GPU execution efficiency while retaining the benefits of the task-parallel approach.

We demonstrated that the underlying combination of methods performs well for accelerating the task computations on hardware designed for heterogeneous computing with state-of-the-art CPU-GPU connections. For the offloaded SPH computations, it achieves up to 2.6 to 3.5 times speedup in comparison to CPU-only baselines when comparing the time required to offload and execute tasks on the GPU. On hardware using conventional PCIe connections between the CPU and GPU, the task computations are not accelerated (the achieved speedup was 0.52 in comparison to the CPU-only baseline) as the CPU-GPU data transfers become the bottleneck.

We furthermore assessed performance over a *full simulation*, demonstrating that the solver functioned in a truly heterogeneous manner with CPU computations overlapping with GPU computations. However, we found the GPU accelerated solver to suffer from slightly increased idle time in-between tasks. We postulate this to be an artefact of the packing process, whereby tasks are packed into large sets of leaf cell level computations offloaded in bundles to ensure the GPU computational and communication capabilities are saturated. Still, the GPU accelerated solver provided a 1.8 times speedup in comparison to 72 threads of a Grace CPU for the computation of a time step. The GPU-accelerated SPH solver's energy efficiency was also assessed. It was found that GPU acceleration provided up to $\sim 30\%$ energy savings for tests conducted on the full Grace-Hopper in comparison to the energy consumed by using only the Grace CPU (with all its 72 cores) to execute the simulation.

Schaller et al. (2024) have shown that SWIFT weak scales extremely well up to tens of thousands of cores. The GPU off-loading approach presented in this paper does not add any further inter-node communications. The domain decomposition system also remains unaltered. Therefore, we expect the GPU offloading approach presented in this paper to weak-scale similarly. Multi-node multi-GPU tests presented in Section 4.2 confirm near-perfect strong scaling of GPU-accelerated SWIFT on up to 4 single-GPU nodes with minor deviations from perfect scaling for the tests with 2 and 3 GPUs ($<7\%$) and prefect scaling on 4 GPUs. Scalability on $O(1000)$ GPUs will be confirmed in future work.

Aiming to benefit from the typically much larger CPU memory to be able to fit more sizeable problems on a single compute node, we chose to retain particle data on CPU memory rather than storing them persistently on GPU memory. Instead, particle data is transferred to and from the GPU as needed. However, SWIFT's current particle data memory layout is incompatible with efficient memory transfers to and from the GPU. Hence, in the prototype implementation presented in this work, CPU-side data re-arrangement and packing is required to allow for efficient data transfers and computations. This was shown to constitute the major performance bottleneck in Section 4.2 as it currently takes up to 80% of total offload time.

Optimising this packing process will be the subject of future work, which we will explore through several avenues. Possible optimisations are for example to minimise the total size of data to be packed, as discussed in Section 4.5. Alternately, it may be eliminated altogether by re-structuring the data used in the CPU code to match the form optimised for the GPU at the expense of additional book-keeping and memory movements on the CPU side. Based on the current time the packing process takes, we expect these optimisations could more than double the currently achieved speedups. Furthermore, the CPU-GPU communication size can in principle also be minimised by a factor of 10 by removing currently duplicated data being transferred (see Section 4.5). This would also greatly improve the code performance on PCIe connected GPUs, where it is evident that the communication costs are severely limiting solver performance as shown in Section 3.3.

## ACKNOWLEDGEMENTS

The authors would like to thank EPSRC and the UK's ARCHER2 eCSE GPU grant programme for funding the research in which the solver presented in this paper was developed with grants EP/W026775/1 and ARCHER2-eCSE02-28.

This work made use of the facilities of the N8 Centre of Excellence in Computationally Intensive Research (N8 CIR) provided and funded by the N8 research partnership and EPSRC (Grant No. EP/T022167/1). The Centre is co-ordinated by the Universities of Durham, Manchester and York.

The authors are also grateful to Alastair Basden, who facilitated access to Durham's hardware testbeds which were extremely useful for code development and testing and for numerous discussions which helped crystalise some of the methods presented in this paper.

The authors would like to thank Lee Margetts for facilitating access to the Gemini 2 machine at the University of Manchester.

We also thank Peter Draper, whose inputs and insights were extremely helpful for developing compilation scripts for the GPU accelerated solver. We are grateful to acknowledge very helpful discussions with our colleague in the wider PAX-HPC project Phil Hasnip.

We are also grateful for the reviewer's feedback which led to significant improvements in the quality of this paper.

Finally, we dedicate this paper to the late Richard Bower who was instrumental in developing the research project and securing funding for the research.

## DATA AVAILABILITY

The datasets in this article were derived from a publicly available code repository accessible via `github.com/abouzied-nasar/SWIFT`. The repository includes the initial conditions and experiment setups required to run the tests we present in this paper.

## CONFLICT OF INTEREST STATEMENT

The authors declare no conflict of interest.

## APPENDIX A: GPU OFFLOAD SUMMARY

The GPU-offload algorithm can be approximated by the following simplified pseudo-code for the computation of one time step. The code is parallelised using pthreads such that each CPU thread running the code executes the code independently of other threads. pthread barriers are used for synchronisation when there are no tasks left to execute:

```
Initialise variables and allocate required CPU and GPU memory
Set tasks_packed = 0 //The number of tasks involved in the next GPU offload
Set leaves_packed = 0 //The number of leaf computations included in the next GPU offload
while (tasks available to be executed)
  Take a task from the task queue // this locks the task data
  if (task.subtype == CPU task)
    Execute the task on this CPU thread
    Unlock the task data
    Unlock and enqueue dependant tasks
  else //The task is a GPU task
    Set n_leaf = 0 //The number of leaf computations found from this task
    Set leaf_interactions = {} // Empty array to record leaf interactions.
    while (Recursion through the task cell tree is possible)
      Recurse to find a leaf cell or a pair of leaf cells
      if (found leaf cell with work to be done)
        leaf_interactions[n_leaf] = this leaf level interaction computation
        n_leaf += 1
    if (no more tasks of the current task.subtype left in this threads queue)
      Set launch_leftovers = true
    tasks_packed += 1
    Set task.n_packed =  0 // number of this task's leaf computations packed into CPU buffers
    while (task.n_packed < n_leaf)
      Pack leaf_interactions[n_packed] into part_send //CPU-side buffer array
      task.n_packed += 1
      leaves_packed += 1 // how many leaf computations are packed into CPU buffers
      if (leaves_packed == Sp) //Ready to offload
        Set launch = true
      if (launch) or (launch_leftovers and task.n_packed == n_leaf)
        Unlock task data
        n_bundles = leaves_packed/Sb // Sb = number of leaf computations per bundle
        for (b = 0; b < nbundles; b++) //Offload data to GPU in bundles of Sp/Sb particles
          Compute start and end index of data in CPU buffer arrays of bundle b
          //All instructions below within this loop are asynchronous
          Copy part_send data of bundle b from CPU buffer array to GPU via stream b
          Launch task computation kernel via stream b
          Copy part_recv data of bundle b from GPU to CPU buffer array via stream b
          Insert a cuda event in stream b //records completion signal of prev. GPU instruction
        for(b = 0; b < nbundles; b++)
          Call cudaEventSynchronize(event of stream b) // wait for completion signal
        for(t = 0; t < tasks_packed; t++)
          lock the task t data // spins until lock is obtained
          for(lc = 0; lc < task.n_packed; lc++)
            Unpack computation lc //Copy results from buffers to CPU particle data
          if (not ((t == tasks_packed - 1) and task.n_packed < n_leaf)
            // Only unlock task data/dependencies if no more leafs to be packed for last one
            Unlock task t data
            Unlock and enqueue tasks dependant on task t
        //Reset packing counters and buffer arrays after launch
        Set leaves_packed = 0
        if (task.n_packed == n_leaf) // No more leafs of this task left to pack
          Set tasks_packed = 0
        else // We still have work to do
          Set tasks_packed = 1 // we'll continue working on this task
```